\documentclass[12pt, fullpage]{article}

\usepackage{amsmath,amssymb,amsfonts}
\usepackage{graphicx}
\usepackage{geometry}
\usepackage{cite}
\usepackage{xcolor}
\usepackage{soul}

\title{\noindent Two-opinions-dynamics generated by inflexibles and non-contrarian and contrarian floaters}

\begin{document}

\baselineskip 20 pt

\noindent {\bf Two-opinions-dynamics generated by inflexibles and non-contrarian and contrarian floaters}
\\\\
F. Jacobs
\\\\
{\em Institute of Biology
\\ 
Leiden University
\\ 
Sylviusweg 72
\\ 
NL-2333 BE Leiden
\\ 
The Netherlands
\\\\
E-mail address: f.j.a.jacobs@biology.leidenuniv.nl}
\\\\
S. Galam
\\\\
{\em CEVIPOF - Centre for Political Research
\\
Sciences Po and CNRS
\\
98, rue de l'Universit\'e
\\
75007 Paris
\\
France
\\\\
E-mail address: serge.galam@sciencespo.fr}
\\\\
Author for correspondence: F. Jacobs
\\\\
{\bf Running title:} Two opinions dynamics
\\\\
{\bf Acknowledgements} F. Jacobs appreciates the support of the research underlying this paper by COST grant COST-STSM-P10-01215.
\\\\
The authors are grateful to the anonymous reviewers and the responsible editor for helpful suggestions which improved the article. 

\maketitle

\newpage

\begin{abstract} \baselineskip 20pt \noindent We assume a community whose members adopt one of two opinions $A$ or $B$. Each member appears as an inflexible, or as a non-contrarian or contrarian floater. An inflexible sticks to its opinion, whereas a floater may change into a floater of the alternative opinion. The occurrence of this change is governed by the local majority rule: members meet in groups of a fixed size, and a floater then changes its opinion provided it is a minority in the group. Subsequently, a non-contrarian floater keeps the opinion as adopted under the local majority rule, whereas a contrarian floater adopts the alternative opinion. Whereas the effects of on the one hand inflexibles and on the other hand non-contrarians and contrarians have previously been studied seperately, the current approach allows us to gain insight in the effect of their combined presence in a community. Given fixed proportions of inflexibles $(\alpha_{A}, \alpha_{B})$ for the two opinions, and fixed fractions of contrarians $(\gamma_{A}, \gamma_{B})$ among the $A$ and $B$ floaters, we derive the update equation $p_{t+1}$ for the overall support for opinion $A$ at time $t+1$, given $p_{t}$. The update equation is derived respectively for local group sizes 1, 2 and 3. The associated dynamics generated by repeated local updates is then determined to identify its asymptotic steady configuration. The full opinion flow diagram is thus obtained, showing conditions in terms of the parameters for each opinion to eventually win the competing dynamics. Various dynamical scenarios are thus exhibited, and it is derived that relatively small densities of inflexibles allow for more variation in the qualitative outcome of the dynamics than higher densities of inflexibles.
\end{abstract}

\\\\

\section{Introduction}

\noindent Within the growing field of sociophysics (see \cite{strike} for the defining paper and \cite{book}, \cite{perony}, \cite{schweitzer} for an impression of the state of the art), a great deal of work has been devoted to opinion dynamics  \cite{castellano}. The seminal Galam models of opinion dynamics  \cite{mino1,mino2} and their unification \cite{uni} play a guiding role in analyzing the process of opinion spreading in communities and in providing possible explanations for the outcome of elections. These models are centered around the {\em local majority rule} (l.m.r.), which is applied either in a deterministic or a probabilistic way. In the basic {\em deterministic} case, supporters of the two opinions present in a community are randomly distributed over groups of a fixed size $L$. Within each group members adopt the opinion that has the majority in that group, after which all group members are recollected again. In case there is no majority in a group, its members stick to their own opinion (i.e., {\em neutral} treatment; the {\em probabilistic} treatment in case of a tie assigns opinions to the group members according to a certain probability distribution). Repeated application of this principle generates what is called {\em randomly localized dynamics with a local majority rule}. In the basic {\em probabilistic} case, the community members are divided among groups of various sizes according to some probability distribution, and within each group all members adopt one of the possible opinions with either certainty (majority rule) or  probability (at a tie in even-sized groups) \cite{mino1} . 

\noindent In the basic deterministic two states opinion model, fast dynamics occurs in which the opinion that originally has the majority eventually will obtain complete presence at the cost of the alternative opinion. In the probabilistic two states opinion model, the final outcome depends on the probability distributions for group sizes and local adaptation. Eventually the state of the community can be either one in which only the opinion with initial majority or minority remains, or one with a perfect consensus on both opinions (see \cite{uni}, which unifies basic probabilistic two states opinion models). 

\noindent In \cite{three} a three states opinion model is introduced in which the community members are randomly distributed over groups of size 3. Within each group the l.m.r. is applied, with the additional rule that in case of a tie all members of the group adopt one of the three opinions according to some probability distribution. It is shown that the dynamics quickly converges to a state in which only one of the three opinions is present, which may be an opinion that initially has a minor presence in the community. In addition, the effect of non-voting persons (abstention, sickness, apathy) was shown to have drastic effect on the asymmetry of the threshold value to power \cite{wonc}.

\noindent As a next step to gain a better insight into opinion dynamics, in \cite{contra1} the basic deterministic two states Galam opinion model is extended by the introduction of so-called contrarians. A {\em contrarian} is a community member who, instead of keeping the opinion it adopted under the l.m.r., switches to the alternative opinion. Contrarian behaviour can manifest itself in various ways, e.g. in adolescents as a strive for individualization, especially in an environment of inflexible opinion supporters (see below), as an expression of conformity with the minority, and as negative voting in order to diminish the support for a majority. Depending on the density\footnote{All opinion dynamics models considered in this article are understood to refer to large communities and subcommunities (e.g. contrarians) in which the size of a subcommunity can effectively be described by its density (the part of the subcommunity's size with respect to the whole community) instead of by discrete whole numbers.} of contrarians as well as on group size, their presence either leads to a stabilization of the opinion dynamics in which one opinion (the one with the lower density of contrarians) dominates the other, to an equilibrium in which neither opinion dominates (in case both opinions have equal densities of contrarians), or (in case of relatively large densities of contrarians for both opinions) to a dynamics in which the dominating opinion constantly alternates between the two opinions. The incorporation of contrarians in opinion dynamics models was a step towards a possible explanation of the ``hung elections'' outcome in the U.S. presidential elections in 2000. Although introducing contrarians to explain ``hung elections" at the time may have been a bit speculative (and being aware that possible other influences such as finite population sizes and exogenous factors influencing opinion dynamics have not been considered), it was concluded that if the assumption was sound, under similar conditions the phenomenon should repeat itself in the following years in democratic countries. And indeed, ``hung elections' occurred again several times as with the German elections in 2002 and 2005 as well as the 2006 Italian elections \cite{contra2}. The origin of contrarian behaviour as well as its implications have been the focus of numerous studies \cite{schneider,stauf,wio1,wio2,mobi1,chiche,masuda,kasia1,weis,chatt,taksu,nuno1,timpa, mas1,mas2,moham}.
 
\noindent In addition to the incorporation of contrarian behaviour, the basic deterministic two states Galam model has been modified introducing opinion supporters that express what in politics (and other games) is called {\em inflexible behaviour} \cite{mosco,inflex1}. An inflexible community member is a supporter that under all conditions sticks to its opinion. Under this terminology supporters that switch opinion when in the minority then classify as {\em floaters}, and we shall use this distinction in what follows. In~\cite{inflex1} the effect of inflexible behaviour on opinion dynamics is studied for the case that opinion supporters repeatedly meet in groups of fixed size 3. It is shown that a small density of inflexibles for only one of the two opinions allows for the existence of two local attractors. One of these local attractors is a mixed one, on which both opinions are present and on which the opinion that is supported by inflexibles is a minority. The other attractor is a single state attractor, on which the opinion that is supported by inflexibles has complete majority, i.e., its density equals 1, the other opinion being absent. Due to the presence of these two attractors, the outcome of the opinion dynamics thus depends on the initial condition, the basin of attraction for the mixed local attractor being relatively small compared to that for the single state attractor. If the density of inflexibles is sufficiently large (approximately 17$\%$), the mixed attractor disappears and the single state attractor becomes global. In case both opinions have small and equal densities of inflexibles there are two mixed local attractors. These two attractors are symmetrically situated with regard to a separator on which both opinions are present with density $0.5$. 
 
\noindent A change in the density of inflexibles for one of the opinions breaks this symmetry, and a sufficiently large increase may lead to a global attractor on which the opinion with the larger density of inflexibles has the majority \cite{inflex1}. The inflexible effect could provide for some counter-intuitive explanation to real paradoxical situations \cite{public}. The effect of inflexibles and floaters on opinion dynamics has also been studied extensively in recent years, as seen in \cite{iglesia,martins,mobi2,bolek1,rand,bollen,latora,anten,sob,kasia2,nuno2,lee}.
 
\noindent In this paper we combine the approaches presented in \cite{contra1} and \cite{inflex1}, by allowing for groups composed of inflexibles as well as contrarian and non-contrarian opinion supporters. For clarity we restrict ourselves to groups of fixed size 1,2 and 3. For both opinions we assume fixed densities for the inflexibles. Also, we consider the contrarians to be part of the floaters, i.e., in a given group the contrarians first determine their opinion according to the l.m.r., and subsequently change to become a floater (not necessarily a contrarian) for the alternative opinion (which thus may be the opinion that the contrarian initially was supporting). The presence of contrarians for each opinion is quantitatively expressed as a fixed fraction of the density of floaters of the respective opinion. In case of a tie in groups of size 2 we apply the neutral treatment. After an opinion update, all supporters for both opinions are recollected and then are redistributed again, either as an inflexible or as a non-contrarian or contrarian floater, according to the fixed densities for inflexibles and the fixed fractions of contrarians for the two opinions. We study qualitative characteristics of the opinion dynamics generated by repeated updates. In particular we study changes in the number of equilibria, and changes from monotone to alternating dynamics, due to changes in parameter combinations. The opinion dynamics thus obtained reflects the behaviour of the support for opinions as it is influenced by individuals that for various (e.g. psychological, political) reasons go against the grain as they find themselves in a background consisting of individuals with a clear conviction. A detailed mathematical extension to groups of size 4 will be given in a forthcoming paper \cite{taksu2}. 
\\\\
\noindent{\em Notation}

\noindent We denote the two opinions by $A$ and $B$. The densities of inflexibles for the $A$ and $B$ opinion are denoted by $\alpha_{A}$ and $\alpha_{B}$ respectively, with $0\leq\alpha_{A}\leq 1$ as well as $0\leq\alpha_{B}\leq 1$, and in addition $0\leq\alpha_{A}+\alpha_{B}\leq 1$. Since the roles of the $A$ and $B$ opinion are interchangeable in deriving the opinion dynamics, we may without loss of generality assume that $0\leq\alpha_{A}\leq0.5$, and we shall do so in what follows. The fraction of contrarians among the $A$ floaters is denoted by $\gamma_{A}$, and $\gamma_{B}$ denotes the fraction of contrarians among the $B$ floaters, with both $0\leq\gamma_{A}\leq1$ and $0\leq\gamma_{B}\leq 1$. The size of the groups in which opinion supporters meet is denoted by $L$. The density of the $A$ opinion at time $t=0, 1, 2, \cdots$ (or after $t$ updates) shall be denoted as $p_{t}$. Note that for given $\alpha_{A}$ and $\alpha_{B}$ the density $p_{t}$ necessarily lies in the interval $[\alpha_{A},1-\alpha_{B}]$ (independent of $L$, $\gamma_{A}$ or $\gamma_{B}$). With $f_{L; \alpha_{A}, \alpha_{B}; \gamma_{A}, \gamma_{B}}$ we denote the function that determines the density of the $A$ opinion after application of the l.m.r. followed by the switch of the contrarians. Thus, $p_{t+1}=f_{L;\alpha_{A}, \alpha_{B}; \gamma_{A}, \gamma_{B}}(p_{t})$. Setting $\gamma_{A}=\gamma_{B}=0$, $p_{t+1}=f_{L;\alpha_{A}, \alpha_{B}; 0, 0}(p_{t})$ then gives the density obtained from $p_{t}$ when the l.m.r. is applied without being followed by the switch of the contrarians. In the Appendix tables are given, presenting all possible group compositions in terms of inflexibles and non-contrarian and contrarian floaters for group sizes $L=1$ to 3, together with the effects of the l.m.r. and the opinion changes of contrarians. It is assumed that the community is sufficiently large and well-mixed to allow for the derivation of the density of each possible group composition in the ensemble of all groups of a fixed size from the densities in the community of the constituents of a group. From these tables the expressions for $f_{L;\alpha_{A}, \alpha_{B}; \gamma_{A}, \gamma_{B}}$ are obtained.
\\
\noindent With $\overrightarrow{f_{L; \alpha_{A}, \alpha_{B}; \gamma_{A}, \gamma_{B}}}$ we denote the dynamics generated by repeated application of $f_{L;\alpha_{A},\alpha_{B};\gamma_{A}, \gamma_{B}}$ in subsequent timesteps. Furthermore, $\hat{p}_{L;\alpha_{A}, \alpha_{B};\gamma_{A},\gamma_{B}}$ denotes an asymptotically stable equilibrium for $\overrightarrow{f_{L;\alpha_{A}, \alpha_{B}; \gamma_{A}, \gamma_{B}}}$, and 
$p^{*}_{L;\alpha_{A}, \alpha_{B};\gamma_{A},\gamma_{B}}$ refers to an asymptotically stable periodic point. 
\\
\noindent We now turn to the treatment of the opinion dynamics for group sizes $L=1, 2$ and $3$. 

\section{$L=1$}

\noindent The case $L=1$ resembles a community in which each member is unaffected by other community members in determining its opinion, and the only changes in opinion come from the contrarians. The contributions to the $A$ density after application of the local majority rule is obtained from the second column in Table~\ref{cap:tab1} in Appendix~\ref{subsection:T1}. This column obviously is equal to the first one, since in groups of size $1$ local majority is automatically obtained, but is without effect on the opinion densities. These contributions are: $\alpha_{A}$ for the $A$ inflexibles, and $p-\alpha_{A}$ for the (non-contrarian and contrarian) $A$ floaters. Their sum is $p$, and we obtain for the update rule of the local majority rule that 
\begin{equation}
p_{t+1}=f_{1;\alpha_{A},\alpha_{B};0,0}(p_{t})=p_{t};
\end{equation} 

\noindent consequently, each $p\in[\alpha_{A},1-\alpha_{B}]$ is a neutrally stable equilibrium for the opinion dynamics generated by the l.m.r.. 

\noindent In case only (non-contrarian and contrarian) floaters are involved both $\alpha_{A}$ and $\alpha_{B}$ are equal to 0, and we restrict ourselves to the contributions from the second, third, fifth and sixth line in the table. Since the l.m.r. leaves each group of size 1 unaffected, a switch by a contrarian in this case necessarily implies a change to the opinion it initially does not support. Thus, here also a contribution to the $A$ density comes from the group that initially consists of only $B$ contrarians, as these will turn into $A$ floaters. In this case we obtain for the contribution to the $A$ density: 
\begin{equation}
p_{t+1}=f_{1; 0, 0;\gamma_{A}, \gamma_{B}}(p_{t})=(1-\gamma_{A})p_{t}+\gamma_{B}(1-p_{t})=\gamma_{B}+\Big(1-(\gamma_{A}+\gamma_{B})\Big)p_{t}.
\end{equation} 

\noindent The effect of both inflexibles and non-contrarian as well as contrarian floaters is obtained by adding all the expressions in the last column: the contributions $\alpha_{A}$ due to the invariant density of $A$ inflexibles, $(1-\gamma_{A})(p_{t}-\alpha_{A})$ from the non-contrarian $A$ floaters, and  $\gamma_{B}(1-\alpha_{B}-p_{t})$ from the contrarian $B$ floaters. This yields:

\begin{equation}\label{eq:L=1general}
\begin{tabular}{lll}
$p_{t+1}$ & = & $f_{1;\alpha_{A},\alpha_{B};\gamma_{A},\gamma_{B}}(p_{t})$
\\\\
& = & $\alpha_{A}+(1-\gamma_{A})(p_{t}-\alpha_{A})+\gamma_{B}(1-\alpha_{B}-p_{t})$
\\\\
& = & $\alpha_{A}\gamma_{A}+(1-\alpha_{B})\gamma_{B}+\Big(1-(\gamma_{A}+\gamma_{B})\Big)p_{t}.$ 
\end{tabular}
\end{equation}

\begin{figure}[ht!]
\includegraphics[width=\textwidth, trim=0cm 8cm 0cm 12cm]{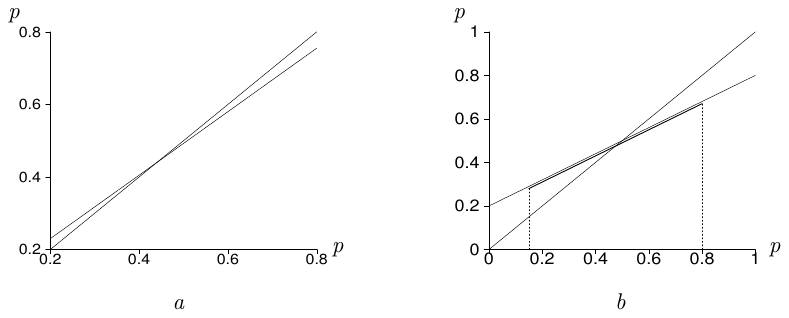}
\caption[Examples for group size $L=1$]{\label{cap:cap1}Figure~{\em a} shows the graphs of $f_{1;0.2, 0.2; 0, 0}$ (which coincides with the diagonal) and $f_{1;0.2, 0.2; 0.075,0.05}$ as functions of $p$ on the interval $[0.2, 0.8]$. By changing the frequencies of contrarians from $(\gamma_{A},\gamma_{B})=(0,0)$ into $(\gamma_{A},\gamma_{B})=(0.075,0.05)$, the collection of neutrally stable equilibria (the diagonal) bifurcates into a unique stable equilibrium $\hat{p}=0.44$ on which the $B$ opinion has the majority. Figure~{\em b} shows the diagonal together with the graph of $f_{1;0, 0; 0.2, 0.2}$ on $[0,1]$, and the graph of $f_{1;0.15, 0.2; 0.2, 0.2}$ on $[0.15, 0.8]$, both as functions of $p$. The dashed lines indicate the boundaries of the interval $[0.15, 0.8]$. The graphs of the two functions almost coincide on this interval and are parallel (due to the equal frequencies of contrarians for both cases). The dynamics generated by these two functions have $\hat{p}=0.5$ and $\hat{p}=0.475$ as their respective stable equilibria.}
\end{figure}

\noindent It follows that if $\gamma_{A}+\gamma_{B}>0$, then 
\begin{equation}
\hat{p}=\dfrac{\alpha_{A}\gamma_{A}+(1-\alpha_{B})\gamma_{B}}{\gamma_{A}+\gamma_{B}}
\end{equation}
is the unique equilibrium for the opinion dynamics $\overrightarrow{f_{1;\alpha_{A}, \alpha_{B}; \gamma_{A}, \gamma_{B}}}$. Due to its linearity as a function of $p_{ t}$, expression~(\ref{eq:L=1general}) implies that the dynamical characteristics of this equilibrium are governed solely by the frequencies of the contrarians. The equilibrium is asymptotically stable if and only if $0<\gamma_{A}+\gamma_{B}<2$. For $0<\gamma_{A}+\gamma_{B}<1$ the equilibrium is approached monotonically, with an increase in the $A$ density if and only if its initial value is less then the equilibrium value. For $\gamma_{A}+\gamma_{B}=1$, the function $f_{1;\alpha_{A}, \alpha_{B}; \gamma_{A}, \gamma_{B}}$ is constant and equals $\alpha_{A}\gamma_{A}+(1-\alpha_{B})\gamma_{B}$; the opinion dynamics then reaches its equilibrium in one iteration. For $1<\gamma_{A}+\gamma_{B}<2$, the equilibrium is approached alternately. For $\gamma_{A}+\gamma_{B}=2$, i.e., both $\gamma_{A}=1$ and $\gamma_{B}=1$, the equilibrium equals $0.5(1+\alpha_{A}-\alpha_{B})$ and is neutrally stable; each $p\in[\alpha_{A},1-\alpha_{B}]$ different from $0.5(1+\alpha_{A}-\alpha_{B})$ generates a neutrally stable cycle of length 2.
\\
\noindent On the equilibrium, the $A$ opinion has the majority if and only if the inequality
\begin{equation}\label{eq:L=1majority}
(0.5-\alpha_{A})\gamma_{A}<(0.5-\alpha_{B})\gamma_{B}
\end{equation}
holds. Thus, for an opinion to achieve the majority it is required that it is being supported by a sufficiently large density of inflexibles, and/or a sufficiently small frequency of contrarians among the floaters.
\\
\noindent Given densities $\alpha_{A}$ and $\alpha_{B}$ of inflexibles for the two opinions, a change in the frequencies of contrarians from 0 into small values $\gamma_{A}$ and $\gamma_{B}$ causes the bifurcation from a collection of neutrally stable equilibria for $\overrightarrow{f_{1;\alpha_{A},\alpha_{B};0,0}}$ into a unique stable equilibrium for $\overrightarrow{f_{1;\alpha_{A},\alpha_{B};\gamma_{A},\gamma_{B}}}$. The opinion which has the majority on this equilibrium is determined by inequality~(\ref{eq:L=1majority}). In case $\alpha_{A}=\alpha_{B}=\alpha$, the opinion with the smaller frequency of contrarians obtains the majority. Conversely, given different frequencies $\gamma_{A}$ and $\gamma_{B}$ of contrarian floaters for the two opinions, in the absence of inflexibles the dynamics $\overrightarrow{f_{1;0,0;\gamma_{A},\gamma_{B}}}$ has $\hat{p}=\frac{\gamma_{B}}{\gamma_{A}+\gamma_{B}}$ as its unique stable equilibrium, on which the opinion with the smaller frequency of contrarians has the majority. Fixing sufficiently small densities $\alpha_{A}$ and $\alpha_{B}$ of both opinions as inflexibles, this equilibrium slightly shifts but leaves the majority unaltered. In case $\gamma_{A}=\gamma_{B}$, in the absence of inflexibles the equilibrium $\hat{p}$ equals $0.5$, and the introduction of small densities of inflexibles for both opinions changes this equilibrium into one on which the opinion with the larger density of inflexibles takes the majority. Figure~\ref{cap:cap1} illustrates these conclusions.  

\newpage

\begin{figure}[hb!]
\includegraphics[width=0.8\textwidth, trim=0cm 3cm 0cm 7.5cm]{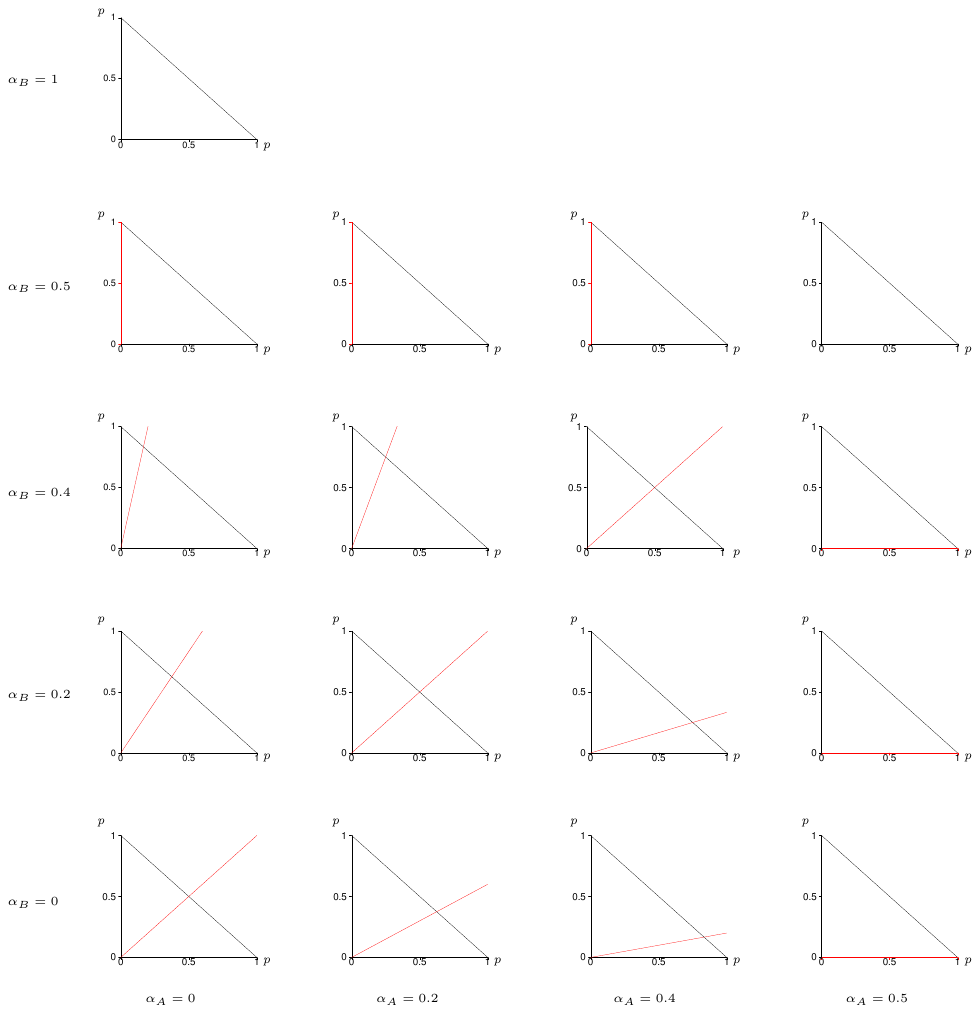}
\caption[Overview of the opinion dynamics for group size $L=1$ ]{\label{cap:L=1overview} An overview of the opinion dynamics of $f_{1; \alpha_{A}, \alpha_{B}; \gamma_{A}, \gamma_{B}}$, for values $\alpha_{A}$ and $\alpha_{B}$ as indicated, and with $\gamma_{A}$ and $\gamma_{B}$ in each pane on the horizontal and vertical axis, respectively, both ranging between 0 and 1. In each pane the line with negative slope $\gamma_{A}+\gamma_{B}=1$ is drawn, and possibly an additional red line of positive (possibly infinite) or zero slope. On the line $\gamma_{A}+\gamma_{B}=1$ the function $f_{1;\alpha_{A}, \alpha_{B};\gamma_{A}, \gamma_{B}}$ is constant, and the corresponding values of $\gamma_{A}$ and $\gamma_{B}$ separate between monotone and alternating dynamics, with the monotone dynamics occurring if $0<\gamma_{A}+\gamma_{B}<1$, i.e., below the line. The red line, if present, gives the values $(\gamma_{1}, \gamma_{2})\neq(0,0)$ for which the equilibrium of the opinion dynamics equals $0.5$, and is determined by the expression $(\alpha_{A}-0.5)\gamma_{A}-(\alpha_{B}-0.5)\gamma_{B}=0$. Opinion $A$ obtains the majority if (and only if) $(\alpha_{A}-0.5)\gamma_{A}-(\alpha_{B}-0.5)\gamma_{B}>0$ holds, i.e., if  $\alpha_{B}<0.5$ and $(\gamma_{A}, \gamma_{B})$ lies above the red line. The panes for values $(\alpha_{A}, \alpha_{B})$ for which $\alpha_{A}+\alpha_{B}=1$ represent degenerate cases, in the sense that only inflexibles for both opinions are present in the community and only one density $\hat{p}=\alpha_{A}$ for the $A$ opinion occurs in time. In case $\alpha_{B}>0.5$, opinion $A$ will never achieve the majority in equilibrium.}
\end{figure}

\noindent Figure~\ref{cap:L=1overview} gives a qualitative overview of the outcomes of the possible opinion dynamics $\overrightarrow{f_{1; \alpha_{A}, \alpha_{B}; \gamma_{A}, \gamma_{B}}}$.

\newpage

\section{$L=2$}

\noindent In groups of size 2 the number of members that support the $A$ or $B$ opinion may be equal, in which case a tie occurs. We shall deal with the neural treatment in case of a tie, in which each supporter keeps its own opinion.
\\
\noindent Table~\ref{cap:tab2} in Appendix~\ref{subsection:T2} is related to groups of size 2. We obtain 
\begin{equation}
p_{t+1}=f_{2; \alpha_{A}, \alpha_{B}; 0, 0}(p_{t})=p_{t},
\end{equation}
which is obvious, since in groups of size 2 no majorities can occur, and, in case of a tie, the neutral application of the local majority rule does not have any effect. Incorporating the effect of non-contrarian as well as contrarian floaters, Table~\ref{cap:tab2} yields that 
\begin{equation}
p_{t+1}=f_{2; \alpha_{A}, \alpha_{B}; \gamma_{A}, \gamma_{B}}(p_{t})=\alpha_{A}\gamma_{A}+(1-\alpha_{B})\gamma_{B}+\Big(1-(\gamma_{A}+\gamma_{B})\Big)p_{t}.
\end{equation}
\noindent Thus, for groups of size 2 the effect of the neutral application of the local majority rule and the contrarians is the same as for groups of size 1. 

\section{$L=3$}

\noindent Group size 3 is the smallest value of $L$ for which the local majority rule becomes effective due to possible group compositions in which a majority of one of the two opinions occurs. As a consequence, the generated dynamics allows for features different from those for group sizes 1 and 2. Careful bookkeeping based on Table~\ref{cap:tab3} in Appendix~\ref{subsection:T3} yields that 
\begin{equation}\label{eq:L=3}
\begin{tabular}{lll}
$p_{t+1}$ & = & $f_{3;\alpha_{A}, \alpha_{B};\gamma_{A}, \gamma_{B}}(p_{t})$
\\\\
& = &  $\alpha_{A}(1 - \gamma_{A}) + (1 - \alpha_{B})\gamma_{B} - \Big(2\alpha_{A}(1 - 2 \gamma_{A}) - \gamma_{A} + \gamma_{B}\Big)p_{t}$ 
\\\\
 & & $+ \Big(3 + \alpha_{A}(1-2\gamma_{A})-\alpha_{B}(1-2\gamma_{B})-4\gamma_{A}-2\gamma_{B}\Big)p_{t}^2 -2\Big(1-\gamma_{A}-\gamma_{B}\Big)p_{t}^3$
\\\\\\
 & = & $p_{t}+\alpha_{A}(1 - \gamma_{A}) + (1 - \alpha_{B})\gamma_{B} - \Big(1 + 2\alpha_{A}(1 - 2 \gamma_{A}) - \gamma_{A} + \gamma_{B}\Big)p_{t}$
\\\\
 & & $+ \Big(3 + \alpha_{A}(1-2\gamma_{A})-\alpha_{B}(1-2\gamma_{B})-4\gamma_{A}-2\gamma_{B}\Big)p_{t}^2 -2\Big(1-\gamma_{A}-\gamma_{B}\Big)p_{t}^3.$
\end{tabular}
\end{equation}
\\
\noindent For clarity we start the analysis of the generated opinion dynamics with the symmetric case of equal densities of inflexibles and equal fractions of contrarians for both opinions.
 
\subsection{The fully symmetric case: $\alpha_{A}=\alpha_{B}$ and $\gamma_{A}=\gamma_{B}$ }\label{subsection:4.1}

\begin{figure}[t!]
\includegraphics[width=\textwidth, trim=0cm 11cm 0cm 9cm]{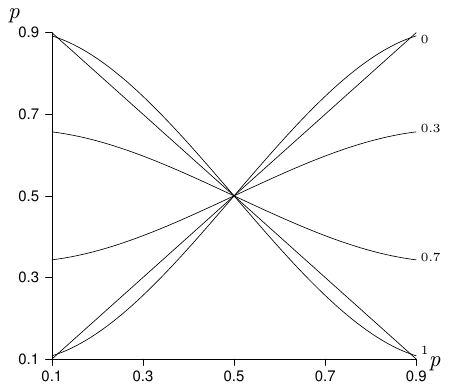}
\caption[Graphs of $f_{3;0.1, 0.1;\gamma, \gamma}$]{\label{cap:cap3alt2}Graphs of $f_{3;0.1,0.1;\gamma, \gamma}$ as function of $p\in[0.1, 0.9]$, with values $\gamma$ as indicated at each specific graph. In addition, the diagonal and the line $1-p$ are drawn.}
\end{figure}

\noindent Taking $\alpha_{A}=\alpha_{B}=\alpha$ and $\gamma_{A}=\gamma_{B}=\gamma$, we obtain that  
\begin{equation}\label{eq:L=3symm}
p_{t+1}=f_{3;\alpha, \alpha; \gamma, \gamma}(p_{t})=p_{t}+(1- 2p_{t})\Big(\gamma+\alpha(1-2\gamma)-(1-2\gamma)p_{t}+(1-2\gamma)p_{t}^{2}\Big)
\end{equation}
\noindent As an illustration to expression~(\ref{eq:L=3symm}), Figure~\ref{cap:cap3alt2} shows a collection of graphs of $f_{3;\alpha, \alpha; \gamma, \gamma}$ as function of $p$, for $\alpha=0.1$ and several values of $\gamma$.

\begin{figure}
\includegraphics[width=\textwidth, trim=0cm 10cm 0cm 12cm]{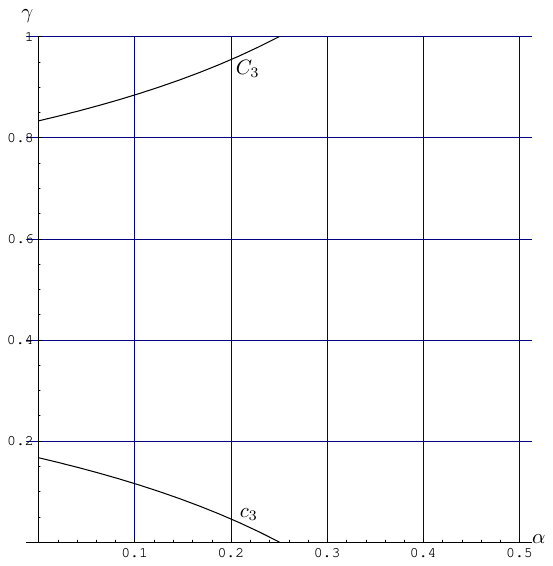}
\caption[Critical curves for the fully symmetrical case]{\label{cap:cap4alt11}The critical curves $c_{3}$ and $C_{3}$ in the $(\alpha,\gamma)$-parameter space}
\end{figure}

\noindent From expression~(\ref{eq:L=3symm}) the analysis of the generated opinion dynamics is straightforward. We give an overview.
\\
\noindent Symmetry considerations imply that the dynamics $\overrightarrow{f_{3;\alpha, \alpha;\gamma, \gamma}}$ has $p=0.5$ as an equilibrium, for any choice of $\alpha\in[0,0.5]$ and $\gamma\in[0,1]$. In addition to parameter combinations $\alpha$ and $\gamma$ for which this equilibrium is unique and stable, there are combinations which allow for an unstable repelling equilibrium $\hat{p}=0.5$ in combination with two other, asymptotically stable, equilibria, or with two asymptotically stable periodic points of minimal period 2. Details for these possibilities to appear are derived in Appendix~\ref{subsection:L=3symm}, here we confine ourselves to the outcome.  
\\
\noindent Let the {\em critical curves} $c_{3}$ and $C_{3}$ be defined as follows: 
\begin{equation}
c_{3}=\{(\alpha,\gamma)\in[0,0.5]\times[0,1]:(3-4\alpha)(1-2\gamma)=2\},
\end{equation}
\noindent and 
\begin{equation}
C_{3}=\{(\alpha,\gamma)\in[0,0.5]\times[0,1]:(3-4\alpha)(1-2\gamma)=-2\}.
\end{equation}
\noindent Figure~\ref{cap:cap4alt11} shows the curves $c_{3}$ and $C_{3}$ in the $(\alpha,\gamma)$-parameter space. On $c_{3}$ the derivative $f'_{3;\alpha, \alpha;\gamma, \gamma}(0.5)$ equals 1, whereas on $C_{3}$ this derivative equals -1. The two corner areas in Figure~\ref{cap:cap4alt11} enclosed by either $c_{3}$ or $C_{3}$ are the regions of parameter combinations for which $0.5$ is unstable; outside these regions (including the curves) $0.5$ is the unique asymptotically stable equilibrium for $\overrightarrow{f_{3;\alpha, \alpha;\gamma, \gamma}}$, independent of the initial condition. The lower left corner region is the area for which the dynamics $\overrightarrow{f_{3;\alpha, \alpha;\gamma, \gamma}}$ has two asymptotically stable equilibria $\hat{p}_{3;\alpha, \alpha;\gamma, \gamma}$. Given parameter combinations $(\alpha,\gamma)$ in this region, the opinion dynamics eventually will stabilize on an equilibrium on which the opinion with the initial majority will have maintained its majority. In case $(\alpha,\gamma)\neq(0,0)$, this equilibrium is mixed; if neither inflexibles nor contrarians are present for both opinions, i.e. $(\alpha,\gamma)=(0,0)$, the equilibrium is a single state attractor with only one opinion present. These results generalize those obtained in \cite{inflex1} for the case of equal densities of inflexibles and no contrarians for both opinions. For parameter combinations in the upper left corner region in Figure~\ref{cap:cap4alt11}, the dynamics has two attracting periodic points of period 2. Here an initial majority does not guarantee the eventual majority, since the dynamics is such that both opinions alternately switch between minority and majority. 

\noindent Thus, if both opinions are being supported by equal densities $\alpha$ of inflexibles and equal fractions $\gamma$ of contrarians among the floaters, for an opinion to obtain the majority it is necessary that $\alpha$ as well as $\gamma$ are sufficiently small, and that it has the initial majority. Also, with increasing $\alpha$ ($\gamma$), the maximum value of $\gamma$ ($\alpha$) for which a majority is attainable decreases. If no inflexibles are present, the fraction of contrarians among the floaters must be less than approximately $17\%$ ($\dfrac{100}{6}\%$) for a majority to be realizable, and if the fraction of contrarians among the floaters equals 0, the density of inflexibles must be less than $25\%$.
\\
If in the parameter space a combination $(\alpha,\gamma)$ approaches from within a corner towards one of the two critical curves, then the two additional equilibria or periodic points approach towards $p=0.5$; a withdrawal in the parameter space results in the opposite movement of the additional equilibria or periodic points. It follows that when passing through $c_{3}$, the dynamics $\overrightarrow{f_{3;\alpha, \alpha;\gamma, \gamma}}$ undergoes a supercritical pitchfork bifurcation, and when passing through $C_{3}$ the dynamics undergoes a period doubling bifurcation (flip bifurcation). 

\subsection{The general case}\label{sect:general}

\begin{figure}
\includegraphics[width=\textwidth, trim=0cm 8.5cm 0cm 4.5cm]{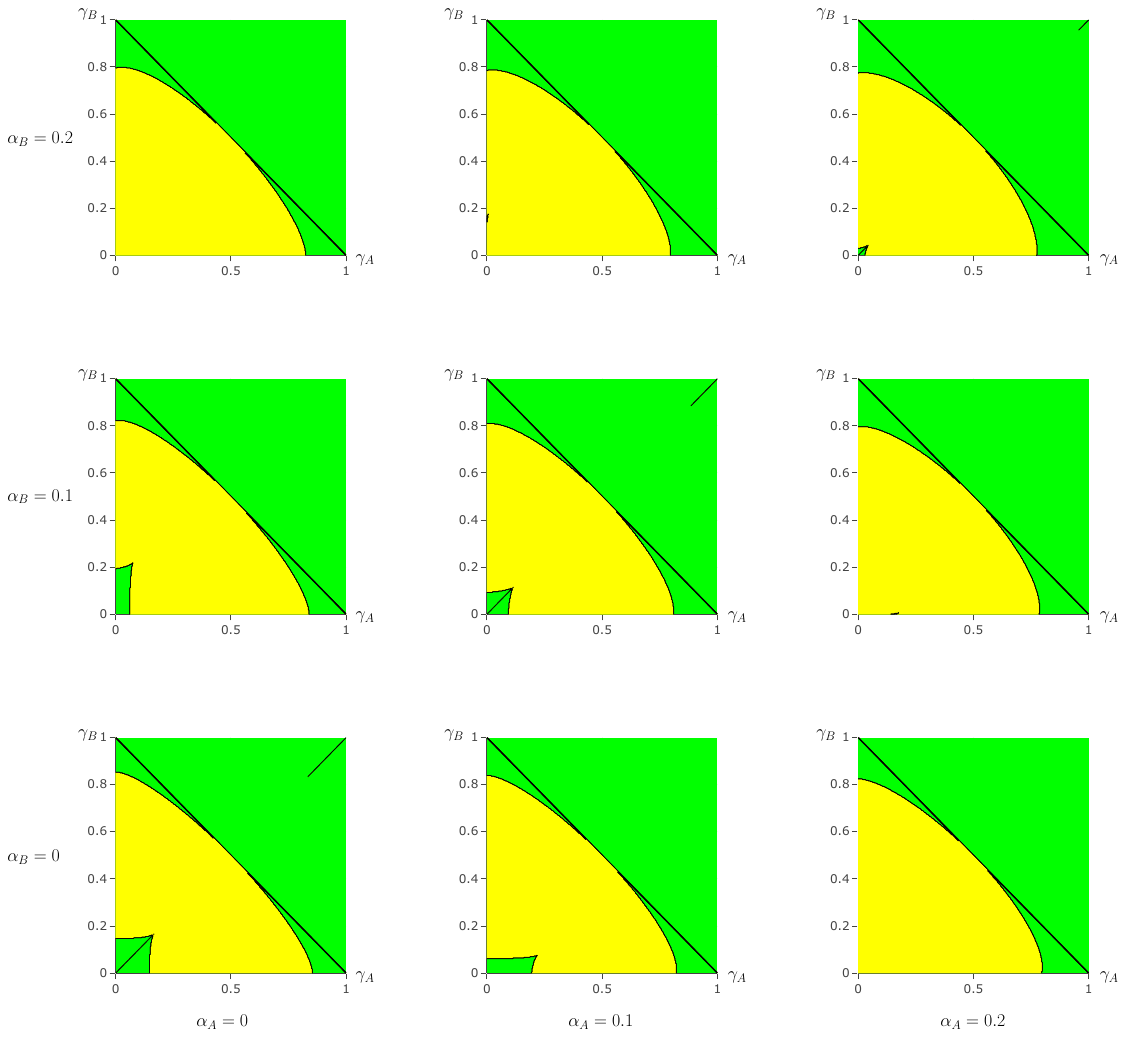}
\caption[Signplots of the discriminant, for small densities of inflexibles for both opinions]{\label{cap:p3selectionalt} A collection of panes, for values $\alpha_{A}$ and $\alpha_{B}$ as indicated, and $\gamma_{A}$ and $\gamma_{B}$ for each pane ranging between 0 and 1, with $\gamma_{A}$ on the horizontal axis and $\gamma_{B}$ on the vertical axis. In each pane the signplot of the discriminant $D(\alpha_{A},\alpha_{B};\gamma_{A},\gamma_{B})$ is shown for points $(\gamma_{A},\gamma_{B})$ for which $\gamma_{A}+\gamma_{B}\neq 1$. Yellow areas represent the parameter combinations with a positive discriminant (i.e., combinations for which the corresponding opinion dynamics has a unique equilibrium), and in green regions the discriminant is negative (the corresponding opinion dynamics then has 3 different equilibria, but not necessarily in the interval $[\alpha_{A}, 1 - \alpha_{B}]$). On the curve separating the yellow and green region the discriminant $D(\alpha_{A}, \alpha_{B}; \gamma_{A}, \gamma_{B})$ for the third-degree function $f_{3; \alpha_{A}, \alpha_{B}; \gamma_{A}, \gamma_{B}}(p)-p$ equals 0 (except in $(\gamma_{A},\gamma_{B})=(0.5, 0.5)$, where this function becomes quadratic). In each pane the line $\gamma_{A}+\gamma_{B}=1$ is drawn in black. On these lines the third-degree function $f_{3; \alpha_{A}, \alpha_{B}; \gamma_{A}, \gamma_{B}}$ becomes quadratic and the corresponding dynamics has a unique equilibrium increasing from 0 (for $\gamma_{A}=1$) to 1 ($\gamma_{A}=0$). Furthermore, in panes for which $\alpha_{A}=\alpha_{B}$ holds, on the line $\gamma_{A}=\gamma_{B}$ in the green regions (i.e., a negative discriminant) in black the points are indicated for which the equilibrium $\hat{p}=0.5$ for $\overrightarrow{f_{3;\alpha_{A},\alpha_{B};\gamma_{A},\gamma_{B}}}$ is unstable; other points on the lines $\gamma_{A}=\gamma_{B}$ (for $\alpha_{A}=\alpha_{B}$) indicate parameter combinations for which $\hat{p}=0.5$ is stable (as follows from Figure~\ref{cap:cap4alt11}).}
\end{figure}

\noindent We now return to the general expression~(\ref{eq:L=3}) and give an overview of the possible outcomes of the dynamics $\overrightarrow{f_{3;\alpha_{A},\alpha_{B};\gamma_{A},\gamma_{B}}}$. The analytical background is given in Appendix~\ref{subsection:L=3general}. We distinguish several cases.
\begin{enumerate}
\item[1.] $\gamma_{A}+\gamma_{B}=1$.
\\
For $\gamma_{A}$ and $\gamma_{B}$ such that $\gamma_{A}+\gamma_{B}=1$, the function $f_{3;\alpha_{A}, \alpha_{B};\gamma_{A}, \gamma_{B}}$ is quadratic in $p$. The corresponding opinion dynamics $\overrightarrow{f_{3;\alpha_{A},\alpha_{B};\gamma_{A}, 1 - \gamma_{A}}}$ has a unique stable equilibrium in the interval $[\alpha_{A}, 1 - \alpha_{B}]$. For $(\gamma_{A},\gamma_{B})=(0.5,0.5)$, the function $f_{3; \alpha_{A},\alpha_{B};0.5, 0.5}$ becomes constant and equals $f_{3; \alpha_{A},\alpha_{B};0.5, 0.5}(p)=0.5(1+\alpha_{A}-\alpha_{B})$; it allows for a unique stable equilibrium $\hat{p}=0.5(1+\alpha_{A}-\alpha_{B})$, on which opinion $A$ has the majority if and only if $\alpha_{A}>\alpha_{B}$. The following figure distinguishes between parameter combinations $\alpha_{A}$, $\alpha_{B}$ and $\gamma_{A}$ for which the $A$ opinion obtains either the majority or minority in equilibrium, and for which the equilibrium is approached monotonically or alternately (Figure~\ref{cap:equimajorswitch}). It follows that with increasing value of $\gamma_{A}$ the region of parameter combinations $(\alpha_A{}, \alpha_{B})$ for which opinion $A$ obtains the majority decreases. In addition, if $\gamma_{A}\leq0.5$, the $A$ opinion can obtain the majority for any value of $\alpha_{A}$, provided that $\alpha_{B}$ is sufficiently small; if $\gamma_{A}>0.5$, $\alpha_{A}$ must be sufficiently large and $\alpha_{B}$ sufficiently small for an $A$ majority to occur. 

\begin{figure}[t]
\includegraphics[width=\textwidth, trim=0cm 7cm 0cm 12cm]{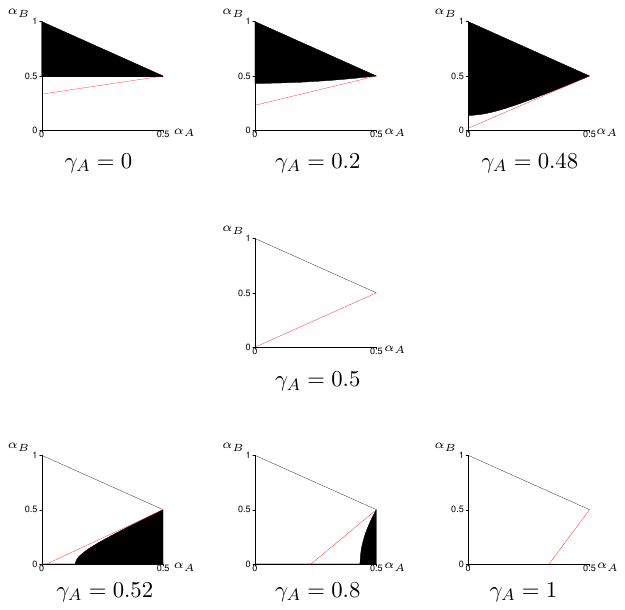}
\caption[Equilibrium densities $\hat{p}_{3;\alpha_{A},\alpha_{B};\gamma_{A}, 1 - \gamma_{A}}$]{\label{cap:equimajorswitch} The different panes, distinguished by different values of $\gamma_{A}$, have  $\alpha_{A}$ on the horizontal axis and $\alpha_{B}$ on the vertical one. Each pane shows the red line $\alpha_{B}=\dfrac{1-2\gamma_{A}}{3-2\gamma_{A}}+\dfrac{1+2\gamma_{A}}{3-2\gamma_{A}}\alpha_{A}$ of parameter values $(\alpha_{A}, \alpha_{B})$ for which the equilibrium value $\hat{p}_{3;\alpha_{A},\alpha_{B};\gamma_{A}, 1 - \gamma_{A}}$ equals $0.5$. Below a red line the equilibrium value lies above $0.5$, i.e., opinion $A$ then obtains the majority. In addition each pane shows in white the region of parameters $(\alpha_{A}, \alpha_{B})$ for which the equilibrium $\hat{p}_{3;\alpha_{A},\alpha_{B};\gamma_{A}, 1 - \gamma_{A}}$ is approached monotonically; the black regions indicate parameter combinations for which the equilibrium is approached alternately. For $\gamma_{A}=0.5$, the derivative of $f_{3;\alpha_{A},\alpha_{B};\gamma_{A}, 1 - \gamma_{A}}$ in the equilibrium equals 0 for all parameter values $(\alpha_{A}, \alpha_{B})$, and the equilibrium is reached in one iteration. On the line $\alpha_{A}+\alpha_{B}=1$ the dynamics is degenerate: the density $p$ is restricted to a single equilibrium density $\hat{p}=\alpha_{A}$. The white region $\alpha_{A}+\alpha_{B}>1$ is not involved in the analysis.}
\end{figure}

\newpage

\item[2.]$\gamma_{A}+\gamma_{B}\neq1$. 
\\
The expression $f_{3;\alpha_{A},\alpha_{B};\gamma_{A},\gamma_{B}}(p)-p=0$ for determining the equilibria is 
\begin{equation}\label{eq:g}
\begin{tabular}{l}
$f_{3;\alpha_{A},\alpha_{B};\gamma_{A},\gamma_{B}}(p)-p$
\\\\
$=\alpha_{A}(1 - \gamma_{A}) + (1 - \alpha_{B})\gamma_{B} - \Big(1 + 2\alpha_{A}(1 - 2 \gamma_{A}) - \gamma_{A} + \gamma_{B}\Big)p+$ 
\\\\
$\Big(3 + \alpha_{A}(1-2\gamma_{A})-\alpha_{B}(1-2\gamma_{B})-4\gamma_{A}-2\gamma_{B}\Big)p^2
-2\Big(1-\gamma_{A}-\gamma_{B}\Big)p^3$
\\\\
$=0.$
\end{tabular}
\end{equation}
The number of solutions is determined by its discriminant, which is denoted by 
\\
$D(\alpha_{A}, \alpha_{B}; \gamma_{A},\gamma_{B})$. The expression for the discriminant is derived in Appendix~\ref{subsection:L=3general}; here we discuss its implications.
\\
For parameter combinations $(\alpha_{A},\alpha_{B};\gamma_{A}, \gamma_{B})$ such that $D(\alpha_{A},\alpha_{B};\gamma_{A}, \gamma_{B})>0$, the equation $f_{3; \alpha_{A},\alpha_{B};\gamma_{A},\gamma_{B}}(p)-p=0$ has a unique real solution. If $D(\alpha_{A},\alpha_{B};\gamma_{A},\gamma_{B})<0$, there are three real solutions. However, these solutions do not necessarily have to belong to the interval $[\alpha_{A}, 1 - \alpha_{B}]$ (but if a solution lies in this interval, it clearly is an equilibrium for the dynamics $\overrightarrow{f_{3;\alpha_{A},\alpha_{B};\gamma_{A},\gamma_{B}}}$). If $D(\alpha_{A},\alpha_{B};\gamma_{A},\gamma_{B})=0$ there are three real solutions, of which at least two coincide; if this happens in the interval $[\alpha_{A}, 1-\alpha_{B}]$, the parameter combination is at a bifurcation point, discriminating between dynamics with either a unique equilibrium or three equilibria. If at the bifurcation point exactly two of the three solutions coincide, the coinciding solutions form a semistable equilibrium. 
\\
Figure~\ref{cap:p3selectionalt} shows a collection of signplots for the discriminant, for values $\alpha_{A}$ and $\alpha_{B}$ as indicated, and with $\gamma_{A}$ and $\gamma_{B}$ for each signplot between 0 and 1. In addition the outcome of the analysis for parameter combinations $(\alpha_{A},\alpha_{B}; \gamma_{A}, 1 - \gamma_{A})$ is included, as well as the results of the analysis for combinations $(\alpha, \alpha; \gamma, \gamma)$. 
\\
The discriminant becomes singular for parameter combinations $(\alpha_{A},\alpha_{B};\gamma_{A},\gamma_{B})$ with $\gamma_{A}+\gamma_{B}=1$. In approaching such parameter combinations for which $(\gamma_{A}, \gamma_{B})\neq(0.5, 0.5)$, the value of $D(\alpha_{A},\alpha_{B};\gamma_{A},\gamma_{B})$ goes to $-\infty$.  For $(\gamma_{A}, \gamma_{B})=(0.5, 0.5)$, the limit generically equals $+\infty$ when this point is approached from the region $\gamma_{A}+\gamma_{B}<1$; the limit equals $-\infty$ in case it is approached from the other side, i.e., from the region $\gamma_{A}+\gamma_{B}>1$. (In case $(0.5, 0.5)$ is approached along the zero set of $D(\alpha_{A}, \alpha_{B}; \gamma_{A}, \gamma_{B})$, i.e., in each pane in Figure~\ref{cap:p3selectionalt} along the boundary that distinguishes between the yellow and green regions and touches with the line $\gamma_{A}+\gamma_{B}=1$, the limit clearly equals 0.) 
\\
Our further discussion of the opinion dynamics $\overrightarrow{f_{3;\alpha_{A},\alpha_{B};\gamma_{A},\gamma_{B}}}$ is based on Figure~\ref{cap:p3selectionalt}. Instead of a detailed analytical treatment, we continue with a number of characteristic outcomes of the opinion dynamics. 
\end{enumerate}

\newpage

\begin{figure}
\includegraphics[width=\textwidth, trim=0cm 3.5cm 0cm 4cm]{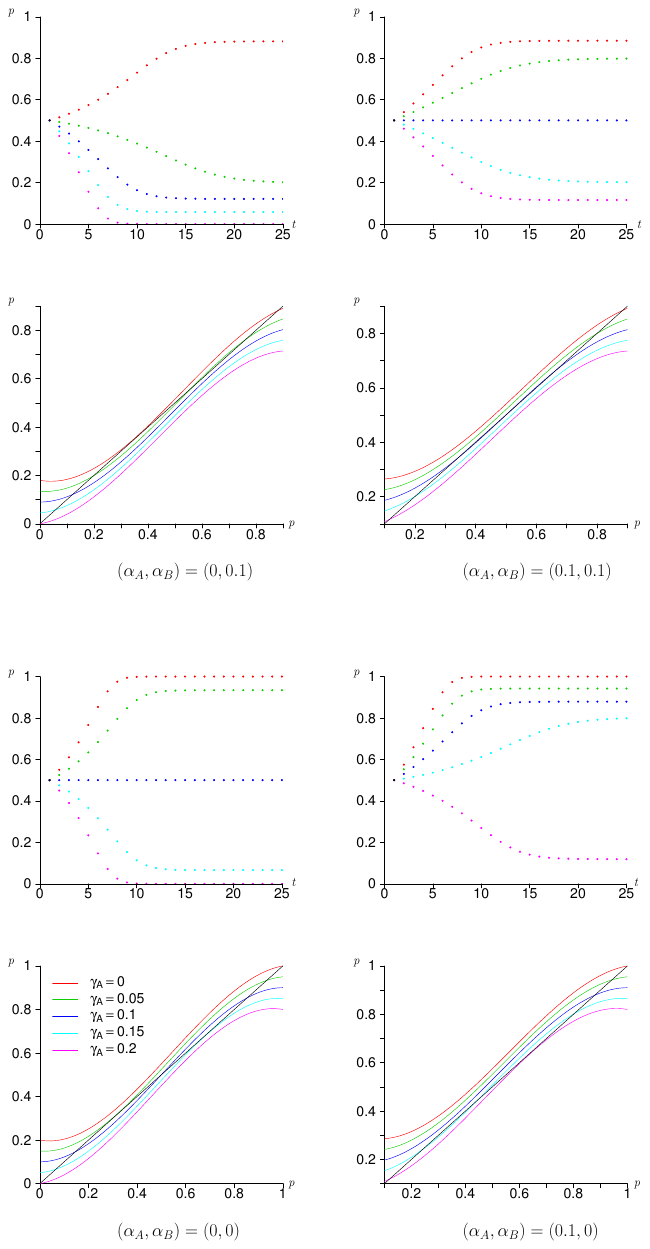}
\caption[Examples of opinion dynamics for small densities of inflexibles and small fractions of contrarians for both opinions]{\label{cap:qallsmall} Four panes of graphs of functions $f_{3;\alpha_{A},\alpha_{B};\gamma_{A}, \gamma_{B}}$ for relatively small values $\alpha_{A}$, $\alpha_{B}$, $\gamma_{A}$ and $\gamma_{B}$, with $(\alpha_{A}, \alpha_{B})$ as indicated below each pane, and with values $\gamma_{A}$ as indicated by the color code. In each of the four panes, $\gamma_{A}$ and $\gamma_{B}$ satisfy $\gamma_{A}+\gamma_{B}=0.2$. I.e., in the corresponding panes in Figure~\ref{cap:p3selectionalt} we traverse the line $\gamma_{A}+\gamma_{B}=0.2$ from its upper left point on the $\gamma_{A}=0$ axis to its lower right point on the $\gamma_{B}=0$ axis, thus passing through regions with positive, zero as well as negative discriminant. Above each of these panes the values of the densities for opinion $A$ in subsequent timesteps are plotted, as obtained by the corresponding opinion dynamics $\overrightarrow{f_{3;\alpha_{A}, \alpha_{B}; \gamma_{A}, \gamma_{B}}}$, with initial density $p=0.5$.}
\end{figure}

\noindent A first characteristic that draws attention in Figure~\ref{cap:p3selectionalt} is the existence of a wedge-shaped region of parameter combinations $(\alpha_{A}, \alpha_{B}; \gamma_{A}, \gamma_{B})$ with negative discriminant for sufficiently small values of all four parameters. For the cases with both $\alpha_{A}=\alpha_{B}$ and $\gamma_{A}=\gamma_{B}$ within this region, we already found the existence of two attracting equilibria, symmetrically positioned with respect to a third, unstable equilibrium $0.5$. We therefore expect also to find a similar pattern of three equilibria in $[\alpha_{A}, 1 - \alpha_{B}]$ for deviations from such symmetric cases within the wedge-shaped region. In~\cite{inflex1} it has been derived that this is indeed the case in the absence of contrarians, i.e., for parameter combinations for which $\gamma_{A}=\gamma_{B}=0$, and for $\alpha_{A}$ and $\alpha_{B}$ sufficiently small. Figure~\ref{cap:qallsmall}, which shows a number of graphs of functions $f_{3; \alpha_{A}, \alpha_{B}; \gamma_{A}, \gamma_{B}}$ for relatively small values $\alpha_{A}$, $\alpha_{B}$, $\gamma_{A}$ and $\gamma_{B}$, implies the same pattern: in case the determinant $D(\alpha_{A}, \alpha_{B}; \gamma_{A}, \gamma_{B})$ is negative, the opinion dynamics has two attracting equilibria that are separated by an unstable one. The two attracting equilibria differ with respect to the opinion by which they are dominated. By leaving the wedge-shaped area, a bifurcation in the opinion dynamics occurs on its boundary $D(\alpha_{A},\alpha_{B};\gamma_{A}, \gamma_{B})=0$. Generically, when moving from inside the wedge-shaped area towards this boundary, the unstable equilibrium and one of the two stable equilibria move towards each other, and at the bifurcation point merge (thus causing a supercritical saddle-node bifurcation). Once the boundary has been crossed, the region of parameters with a positive discriminant is entered, and the dynamics is left with one attracting equilibrium. On this equilibrium opinion $A$ dominates if the upper part of the boundary is crossed, i.e., when $\gamma_{B}>\gamma_{A}$; opinion $B$ has the majority when the right-hand side of the boundary is passed, on which $\gamma_{A}>\gamma_{B}$ holds. This is also illustrated in Figure~\ref{cap:qallsmall}. The occurrence of such a bifurcation may lead to a drastic change in the outcome of the opinion dynamics: whereas inside the wedge-shaped region the outcome of the opinion dynamics depends on the initial condition, outside the wedge-shaped area the opinion dynamics will end on the unique equilibrium, independent of the initial condition. At the bifurcation point at the endpoint of the sharp region of the wedge-shaped area a supercritical pitchfork bifurcation occurs, in which the three equilibria merge together into one attracting equilibrium. 

\begin{figure}
\includegraphics[width=\textwidth, trim=0cm 8cm 0cm 8cm]{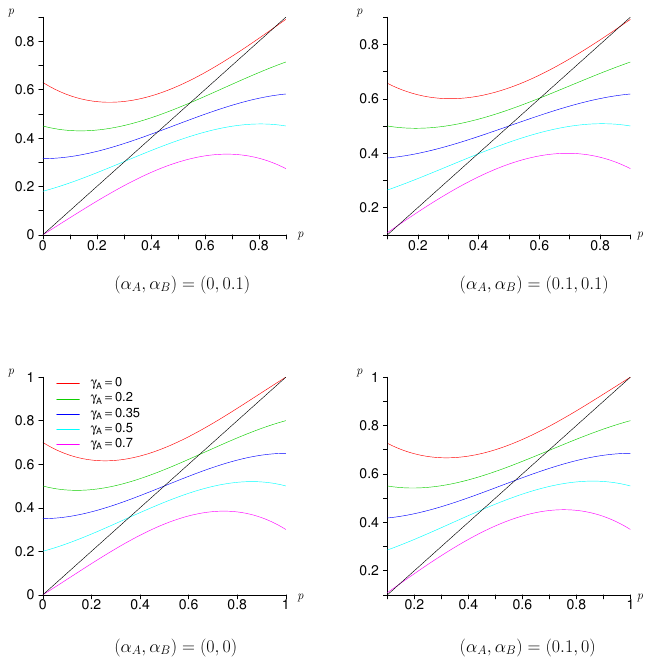}
\caption[Examples of opinion dynamics for small densities for inflexibles for both opinions, and with a positive discriminant $D(\alpha_{A}, \alpha_{B}; \gamma_{A}, \gamma_{B})$]{\label{cap:psmallposdis} Four panes of graphs of functions $f_{3;\alpha_{A},\alpha_{B};\gamma_{A}, \gamma_{B}}$ for relatively small values $\alpha_{A}$ and $\alpha_{B}$ as indicated, and $\gamma_{A}$-values as given by the color code. $\gamma_{A}$ and $\gamma_{B}$  satisfy $\gamma_{A}+\gamma_{B}=0.7$, i.e. for given $(\alpha_{A}, \alpha_{B})$, we traverse the line $\gamma_{A}+\gamma_{B}=0.7$ from its upper left point on the $\gamma_{A}=0$ line to its lower right point on the $\gamma_{B}=0$ line. The discriminant $D(\alpha_{A}, \alpha_{B}; \gamma_{A}, \gamma_{B})$ for the exposed parameter values is positive, indicating a unique equilibrium for the corresponding opinion dynamics.}
\end{figure}

\noindent The yellow regions in Figure~\ref{cap:p3selectionalt} are formed by the parameter combinations for which the discriminant $D(\alpha_{A},\alpha_{B};\gamma_{A},\gamma_{B})$ is positive. The corresponding opinion dynamics then have a unique equilibrium, which (for the parameter combinations in Figure~\ref{cap:p3selectionalt}) is approached monotonically. Figure~\ref{cap:psmallposdis} shows a number of graphs $f_{3;\alpha_{A}, \alpha_{B}; \gamma_{A}, \gamma_{B}}$ for parameter combinations with a positive discriminant. The Figure indicates that for small values of $\gamma_{A}$ and large values of $\gamma_{B}$ opinion $A$ dominates in equilibrium, and that the dominion shift towards the alternative opinion if the fraction of contrarians among the $A$ floaters increases and that among the $B$ floaters decreases.
  
\begin{figure}
\includegraphics[width=\textwidth, trim=0cm 3.5cm 0cm 4cm]{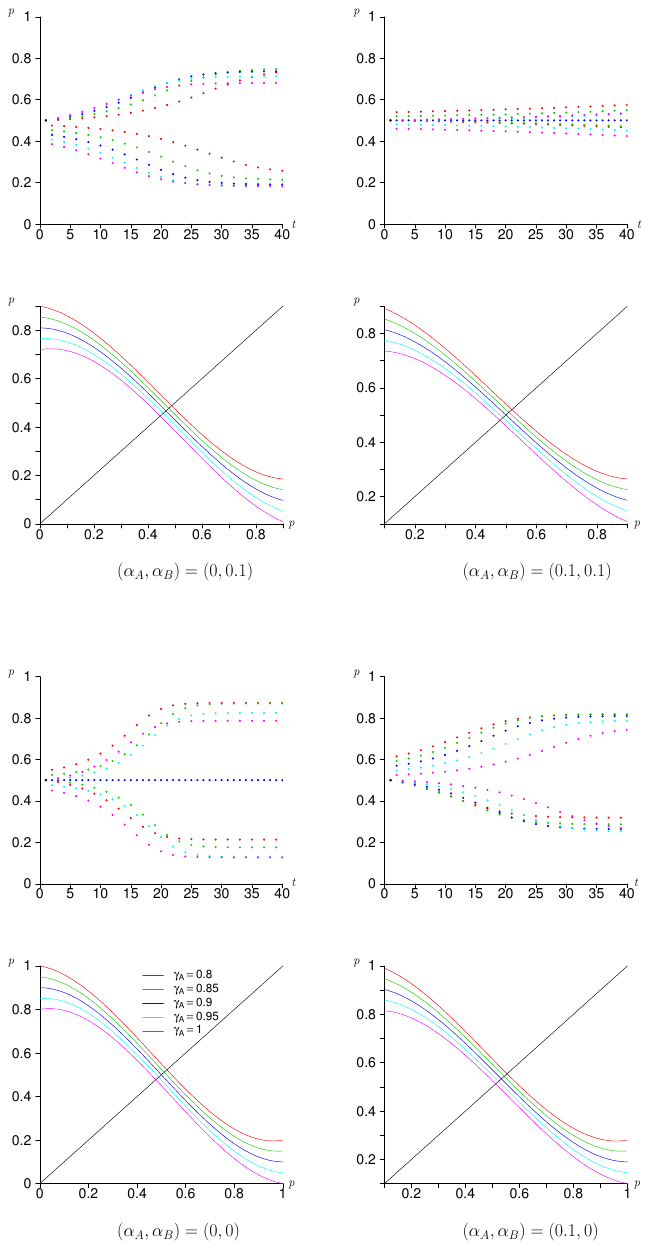}
\caption[Examples of opinion dynamics for small inflexible densities and large fractions of contrarians for both opinions]{\label{cap:psmallupperright} Four panes of graphs of functions $f_{3;\alpha_{A},\alpha_{B};\gamma_{A}, \gamma_{B}}$ for relatively small values $\alpha_{A}$ and $\alpha_{B}$, and with $\gamma_{A}$ and $\gamma_{B}$ satisfying $\gamma_{A}+\gamma_{B}=1.8$. The values of $\alpha_{A}$ and $\alpha_{B}$ are indicated below each of the four panes, and values for $\gamma_{A}$ are as indicated by the color code. I.e., for given $(\alpha_{A}, \alpha_{B})$, we traverse the line $\gamma_{A}+\gamma_{B}=1.8$ from its upper left point on the $\gamma_{B}=1$ line to its lower right point on the $\gamma_{A}=1$ line. Above each of these panes the densities for opinion $A$ are again presented, as obtained by the corresponding opinion dynamics $\overrightarrow{f_{3;\alpha_{A}, \alpha_{B}; \gamma_{A}, \gamma_{B}}}$, with initial density $p=0.5$.}
\end{figure}

\begin{figure}
\includegraphics[width=\textwidth, trim=0cm 3.5cm 0cm 4cm]{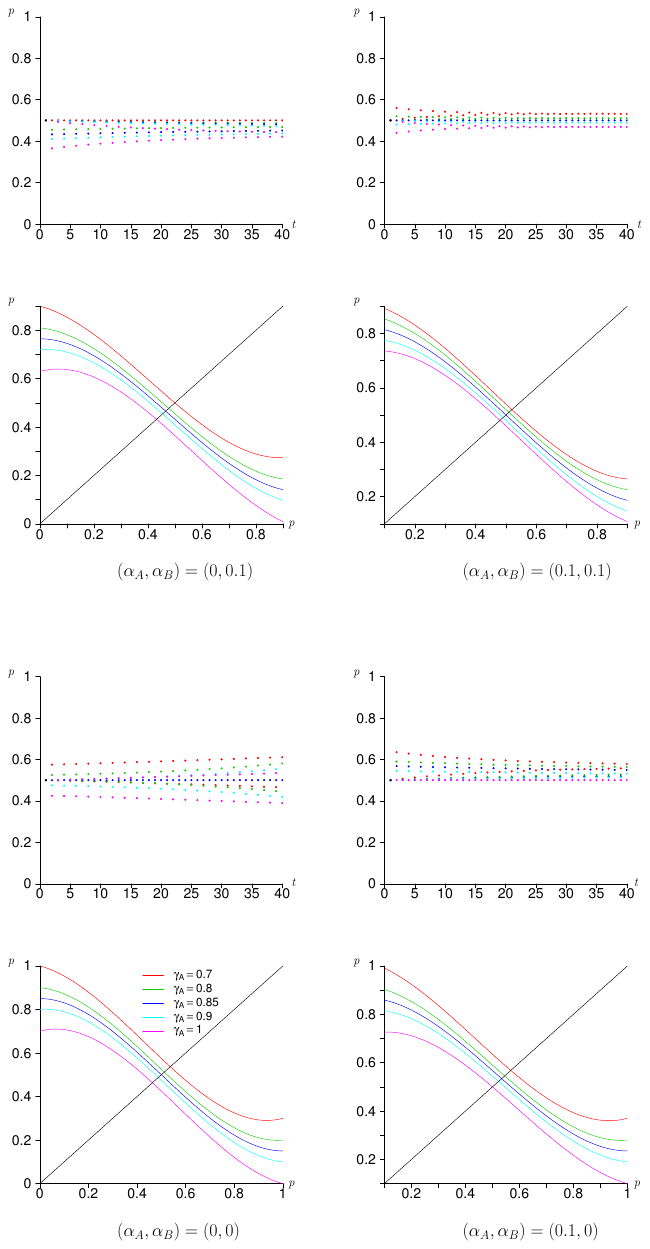}
\caption[Examples of small inflexible densities and large fractions of contrarians that allow for a collapse to equilibrium]{\label{cap:psmallupperrightconv} Four panes of graphs of functions $f_{3;\alpha_{A},\alpha_{B};\gamma_{A}, \gamma_{B}}$ for relatively small values $\alpha_{A}$ and $\alpha_{B}$, and with $\gamma_{A}$ and $\gamma_{B}$ satisfying $\gamma_{A}+\gamma_{B}=1.7$. The values of $\alpha_{A}$ and $\alpha_{B}$ are indicated below each of the four panes, values for $\gamma_{A}$ are again given by the color code. For given $(\alpha_{A}, \alpha_{B})$ values of $\gamma_{A}$ are such that we traverse the line $\gamma_{A}+\gamma_{B}=1.7$ from its upper left point on the $\gamma_{B}=1$ line to its lower right point on the $\gamma_{A}=1$ line. Above each pane the densities for opinion $A$ are presented, as obtained by the corresponding opinion dynamics $\overrightarrow{f_{3;\alpha_{A}, \alpha_{B}; \gamma_{A}, \gamma_{B}}}$, with initial density $p=0.5$.}
\end{figure}

\noindent For given parameters $\alpha_{A}$ and $\alpha_{B}$, crossing the boundary of the yellow area in any direction away from the lower left corner leads to the occurrence of a saddle-node bifurcation, now however outside the domain $[\alpha_{A},1-\alpha_{B}]$ of the functions $f_{3;\alpha_{A},\alpha_{B};\gamma_{A},\gamma_{B}}$ (maintaining an attracting equilibrium in the domain). Therefore in the green region thus entered, the opinion dynamics also is characterized by a unique attracting equilibrium. Proceeding towards the upper right corner, the line of parameter combinations $(\gamma_{A},\gamma_{B})$ satisfying $\gamma_{A}+\gamma_{B}=1$ is crossed. On this line the discriminant $D(\alpha_{A},\alpha_{B};\gamma_{A},\gamma_{B})$ is singular, and the corresponding opinion dynamics have been analyzed in \ref{sect:general}.1. 
\\
In the green area in the upper right corner, for equal and sufficiently small values $\alpha_{A}=\alpha_{B}=\alpha$ and sufficiently large and equal values  $\gamma_{A}=\gamma_{B}=\gamma$ it has been derived earlier that $\overrightarrow{f_{\alpha_{A},\alpha_{B};\gamma_{A},\gamma_{B}}}$ has a unique unstable equilibrium $\hat{p}=0.5$, which causes the convergence of the dynamics towards an attracting periodic orbit of period 2. Neither of the two opinions then achieves the definite majority. The values $\alpha$ and $\gamma$ for which this occurs have been derived in \ref{subsection:4.1}, and are represented in Figure~\ref{cap:p3selectionalt} by black line segments in the upper right corners. For these parameter combinations the discriminant of the equation $f_{3; \alpha_{A}, \alpha_{B}; \gamma_{A}, \gamma_{B}}(p)-p=0$ is negative and thus has three different solutions, of which two are situated outside the domain $[\alpha_{A}, 1 - \alpha_{B}]$. Continuity arguments imply that this behaviour will be maintained for parameter combinations sufficiently close to these line segments. Figure~\ref{cap:psmallupperright} illustrates this. If the parameter combinations are sufficiently far removed from these line segments but $\gamma_{A}$ and $\gamma_{B}$ are still relatively large (i.e., for given $\alpha_{A}$ and $\alpha_{B}$, in the upper right corner), the dynamics will converge alternately to a unique equilibrium. I.e., by moving away from the manifold determined by the constraints $\alpha_{A}=\alpha_{B}$ and $\gamma_{A}=\gamma_{B}$ with large values $\gamma_{A}$ and $\gamma_{B}$, a flip bifurcation occurs in which the attracting periodic 2 orbit collapses to an attracting equilibrium point. This is illustrated by Figure~\ref{cap:psmallupperrightconv}. On the attractor the dominion shifts towards opinion $B$ with increasing $\gamma_{A}$ and decreasing $\gamma_{B}$. 

\newpage

\begin{figure}[h]
\includegraphics[width=\textwidth, trim=0cm 14cm 0cm 6cm]{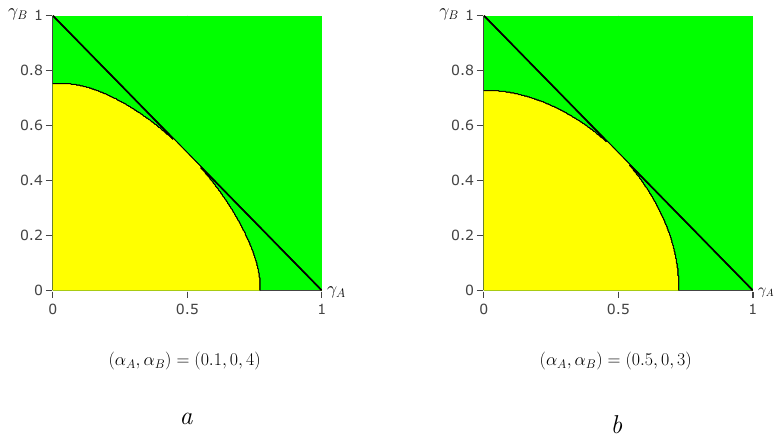}
\caption[Signplots of the discriminants $D(0.1, 0.4; \gamma_{A}, \gamma_{B})$ ({\em a}) and $D(0.5, 0.3; \gamma_{A}, \gamma_{B})$ ({\em b})]{\label{cap:altcases} Signplots of the discriminants $D(0.1, 0.4; \gamma_{A}, \gamma_{B})$ ({\em a}) and $D(0.5, 0.3; \gamma_{A}, \gamma_{B})$ ({\em b}). The color code is as in Fig.~\ref{cap:p3selectionalt}. On the curve separating the yellow and green region the discriminant $D(\alpha_{A}, \alpha_{B}; \gamma_{A}, \gamma_{B})$ for the third-degree function $f_{3; \alpha_{A}, \alpha_{B}; \gamma_{A}, \gamma_{B}}(p)-p$ again equals 0 (except in $(\gamma_{A},\gamma_{B})=(0.5, 0.5)$, where this function becomes quadratic). In addition in each pane the line $\gamma_{A}+\gamma_{B}=1$ is drawn in black.}
\end{figure}

\noindent We end our discussion by presenting some additional opinion dynamics $\overrightarrow{f_{3;\alpha_{A}, \alpha_{B}; \gamma_{A}, \gamma_{B}}}$ for parameter combinations from both the regions with positive and negative discriminant. We remark here that the line segments on the line $\gamma_{A}=\gamma_{B}$ in the lower left and upper right regions of the signplots of $D(\alpha, \alpha; \gamma_{A}, \gamma_{B})$ disappear for $\alpha\geq0.25$. For choices $(\alpha_{A}, \alpha_{B})$ outside the region for which both $\alpha_{A}\leq0.25$ and $\alpha_{B}\leq0.25$ there is no qualitative change in the signplots of $D(\alpha_{A}, \alpha_{B}; \gamma_{A}, \gamma_{B})$, and we choose to restrict and illustrate this for the choices $(\alpha_{A}, \alpha_{B})=(0.1, 0.4)$ and $(\alpha_{A}, \alpha_{B})=(0.5, 0.3)$, i.e., a case with small $\alpha_{A}$ and intermediate $\alpha_{B}$, and one with both $\alpha_{A}$ and $\alpha_{B}$ intermediate. Figure~\ref{cap:altcases} shows the signplots of the discriminants $D(0.1, 0.4; \gamma_{A}, \gamma_{B})$ ({\em a}) and $D(0.5, 0.3; \gamma_{A}, \gamma_{B})$ ({\em b}). 

\begin{figure}
\includegraphics[width=\textwidth, trim=0cm 3cm 0cm 3.5cm]{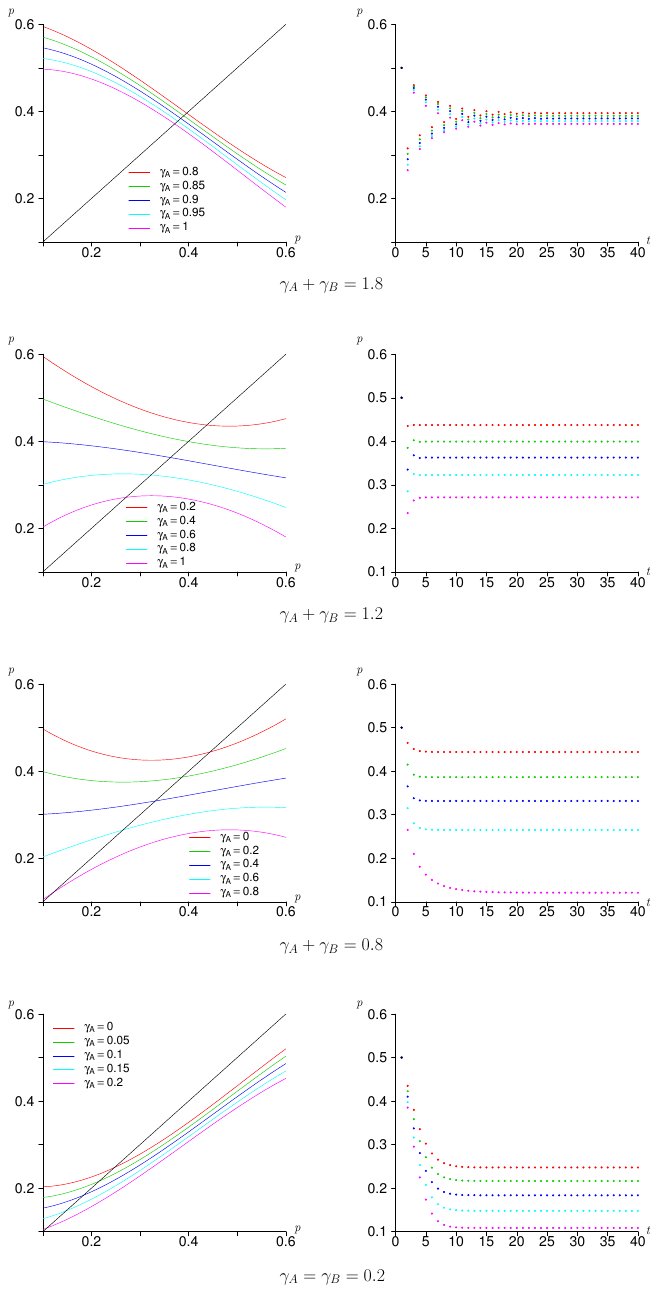}
\caption[Examples of opinion dynamics for small densities of inflexibles for the $A$ opinion and intermediate densities for the $B$ inflexibles]{\label{cap:pasmallbbig} The left column shows four panes of graphs of functions $f_{3;0.1,0.4;\gamma_{A}, \gamma_{B}}$, for different combinations of $\gamma_{A}$ and $\gamma_{B}$, with $\gamma_{A}$ as indicated and per row of graphs $\gamma_{B}$ such that the relation mentioned below the row of graphs is satisfied. The right column of the Figure shows the densities of opinion $A$ as generated by the corresponding opinion dynamics for initial value $p=0.5$.}
\end{figure}

\begin{figure}
\includegraphics[width=\textwidth, trim=0cm 3cm 0cm 3.5cm]{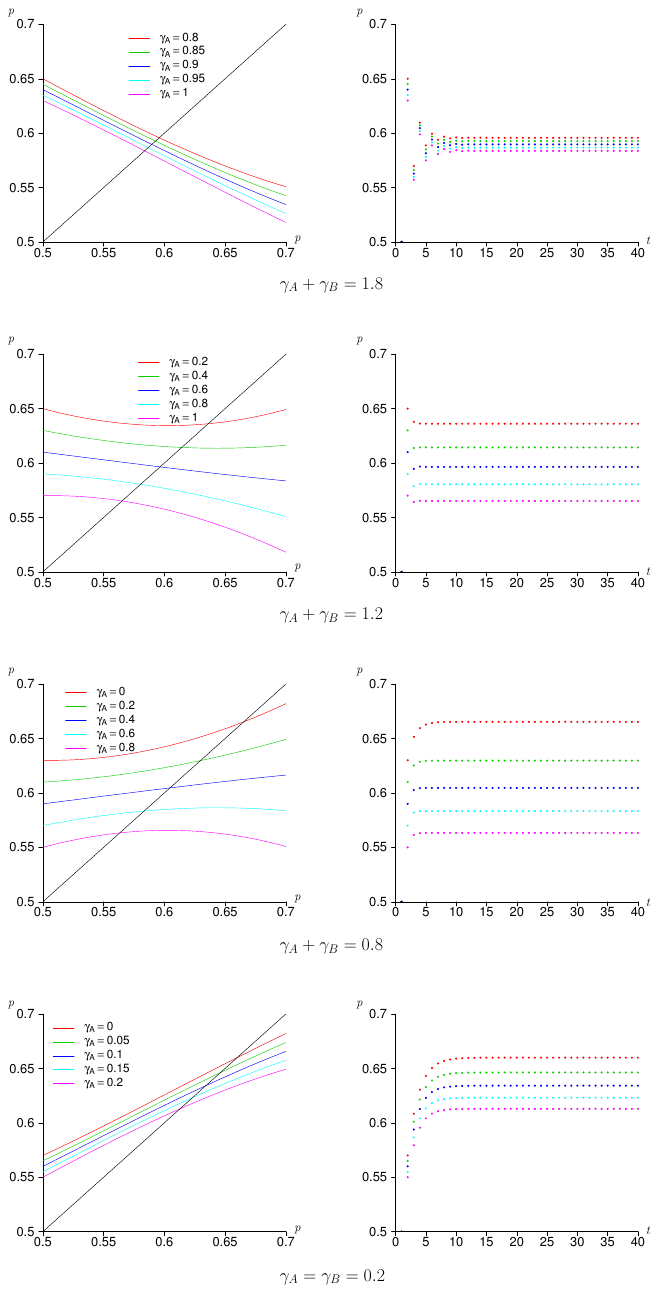}
\caption[Examples of opinion dynamics for intermediate densities of inflexibles for both opinions]{\label{cap:pabintmed} The left column shows four panes of graphs of functions $f_{3;0.5,0.3;\gamma_{A}, \gamma_{B}}$ for the same combinations of $\gamma_{A}$ and $\gamma_{B}$ as in Fig.~\ref{cap:pasmallbbig}. The right column of the Figure shows the densities of opinion $A$ as generated by the corresponding opinion dynamics for initial value $p=0.5$.}
\end{figure}

\noindent The corresponding graphs of $f_{3; \alpha_{A}, \alpha_{B}; \gamma_{A}, \gamma_{B}}$ are represented in Figures~\ref{cap:pasmallbbig} and~\ref{cap:pabintmed}, for several values of $\gamma_{A}$ and $\gamma_{B}$. All cases allow for a unique attracting equilibrium. High values of both $\gamma_{A}$ and $\gamma_{B}$ lead to alternating convergence. Furthermore, a decrease in the fraction of contrarians among the floaters of an opinion increases the density of this opinion in equilibrium. 

\section{Conclusions}

\noindent The results presented re-establish those derived in~\cite{contra1,contra2}, which concerned communities of non-contrarian and contrarian floaters, and~\cite{inflex1}, which studied the combined effects of inflexibles and non-contrarian floaters. The distinctive patterns of opinion dynamics are not characterized by complete quantitative detail. Rather, the results intend to point to possible outcomes of opinion dynamics. We conclude that various kinds of dynamics may occur. In case the local majority rule followed by the contrarian changes are applied for group sizes $L=1$ or 2, the opinion dynamics generically converges to a unique equilibrium. In case the sum of fractions of contrarians for the two opinions is larger than 2, the equilibrium generically is approached alternately, otherwise the dynamics generically shows a monotone approach. For an opinion to obtain the majority in equilibrium, it is required that this opinion is supported by a sufficiently large density of inflexibles in combination with a sufficiently small fraction of contrarians, as expressed by condition~(\ref{eq:L=1majority}).  
\\
Group size $L=3$ allows for additional outcomes for the opinion dynamics. For sufficiently small densities of inflexibles for both opinions, and in addition sufficiently small fractions of contrarians among the floaters for the two opinions, the dynamics allows for two attracting equilibria, that differ in which opinion has the majority. The opinion that eventually will achieve the majority thus is determined by the initial condition, and an opinion that has a fraction of contrarians that is sufficiently smaller than that of the alternative opinion, may achieve the majority in equilibrium, although initially it may be present as a minority. For small values of inflexibles in combination with sufficiently large fractions of contrarians among the floaters, the generated opinion dynamics causes alternating convergence to a period 2 stable orbit (Figure~\ref{cap:psmallupperright}). An increase of the densities of inflexibles or a slight lowering of at least one of the fractions of contrarians causes the collapse of the attracting periodic orbit into an equilibrium, but maintains the alternating behaviour (Figure~\ref{cap:psmallupperrightconv}). For a relatively large collection of parameter combinations the dynamics ends up on a unique attracting equilibrium, which is approached either monotonically or alternating (see e.g. Figures~\ref{cap:equimajorswitch} and~\ref{cap:pabintmed}). An increase in the fraction of contrarians among the floaters of an opinion leads to a decrease of the density of that opinion in equilibrium. Thus, for an opinion to achieve the majority in equilibrium, a small fraction of contrarians among its floaters is favorable.
\\
In \cite{contra2}, the ``hung elections" outcome in several national votes has been discussed in terms of the interplay of non-contrarian and contrarian floaters. Likewise, the present paper may shed a light on the (dis)appearance of alternating opinion dynamics. An alternating series of wins and losses of the majority for two political opinions in pre-election polls may point to considerable fractions of contrarians among the floaters on both sides. In case the alternating pattern converges to a stable period 2 cycle, the uncertainty who will win the election will linger on until the final decisive event. (Note that since the outcome of an election in an alternating environment depends on the moment the election actually takes place, it may happen that in subsequent polls the same winner occurs. This is however no indication of sustained major support. Furthermore, in a sequence of alternating environments, a large number of subsequent wins for the same opinion seems unlikely.) If however the alternating changes are converging to an equilibrium, one of the opinions eventually will reach a decisive majority. Due to the sensitivity of politics for influences, a change in parameter values may easily occur, either with respect to the densities of inflexibles or to the fractions of contrarians. This may result in a switch from the one alternating pattern into the other one, or even into monotone convergence towards an equilibrium. Although our framework does not map unequivocally to real communities, we think it may hint at possible explanations of outcomes of opinion dynamics.
\\
In forthcoming papers we plan to continue the study of opinion dynamics, by focusing on communities in which more than two opinions are being supported, and by taking into account geographic networks.
\newpage

\section{Appendices}

\noindent Subsections~\ref{subsection:T1},~\ref{subsection:T2} and~\ref{subsection:T3} present tables for groups of sizes $L=1, 2$ and 3 from which the density of the $A$ opinion is derived after application of the l.m.r. and the switch by the contrarians, given an initial density $p$ for the $A$ opinion. Each table consists of four columns, of which the first four are separated by arrows. The first column gives the possible group compositions in terms of inflexibles and non-contrarian and/or contrarian floaters, the second column gives the effect of the application of the l.m.r. for the group compositions given in the first column. An application is indicated by a horizontal arrow, whose first appearance in a table is indexed by ``l.m.r."; at other places in the tables this index is omitted. The third column then gives the effect of the switches by the contrarians if applicable, where it is understood that a contrarian switches into a floater of the alternative opinion. The final column gives the contributions of the effect of the l.m.r. and the presence of the contrarians to the density of the $A$ opinion, weighed with the probability of the original group composition in the ensemble of all possible groups of fixed size, given the densities $\alpha_{A}$ and $\alpha_{B}$ of the inflexibles for both opinions, the fractions $\gamma_{A}$ and $\gamma_{B}$ of contrarians among the floaters of the $A$ and $B$ opinion, respectively, and the densities $p-\alpha_{A}$ for the $A$ floaters and $1-\alpha_{B}-p$ for the $B$ floaters. The total sum of these contributions yields $f_{L;\alpha_{A}, \alpha_{B}; \gamma_{A}, \gamma_{B}}(p)$. After each opinion update, all supporters for both opinions are recollected and then redistributed again, either as inflexible or as a non-contrarian or contrarian floater, according to the fixed densities for inflexibles and the fixed fractions of contrarians for the two opinions. 

\noindent In each table the following notation is being used:
\\\\
\noindent\begin{tabular}{ll}
$A_{i}$&inflexible of the $A$ opinion,
\\
$A_{f}$&floater of the $A$ opinion,
\\
$A_{nc}$&non-contrarian floater of the $A$ opinion,
\\
$A_{c}$&contrarian floater of the $A$ opinion,
\\
$A_{f_{nc}}$&floater of the $A$ opinion coming from a $B$ non-contrarian floater after application 
\\
& of the l.m.r.,
\\
$A_{f_{c}}$&floater of the $A$ opinion coming from a $B$ contrarian floater after application 
\\
& of the l.m.r.,
\end{tabular}
\\\\
and
\\
\noindent\begin{tabular}{ll}
$B_{i}$&inflexible of the $B$ opinion,
\\
$B_{f}$&floater of the $B$ opinion,
\\
$B_{nc}$&non-contrarian floater of the $B$ opinion,
\\
$B_{c}$&contrarian floater of the $B$ opinion,
\\
$B_{f_{nc}}$&floater of the $B$ opinion coming from a $A$ non-contrarian floater after application 
\\
& of the l.m.r.,
\\
$B_{f_{c}}$&floater of the $B$ opinion coming from a $A$ contrarian floater after application 
\\
& of the l.m.r..
\end{tabular}

\newpage

\subsection{Table 1: group size $L=1$}\label{subsection:T1}

\begin{table}[ht!]
\caption{}\label{cap:tab1} 
\includegraphics[width=\textwidth, trim=0cm 3cm 0cm 3cm]{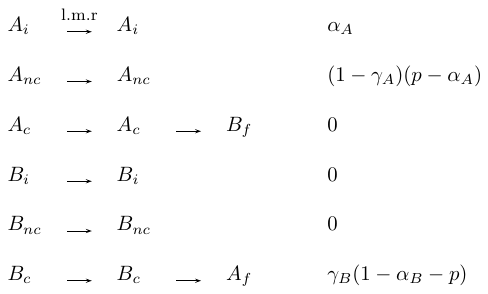}
\end{table}

\newpage

\subsection{Table 2: group size $L=2$}\label{subsection:T2}

\begin{table}[ht!]
\caption{}\label{cap:tab2} 
\includegraphics[width=0.8\textwidth, trim=0cm 3cm 0cm 3cm]{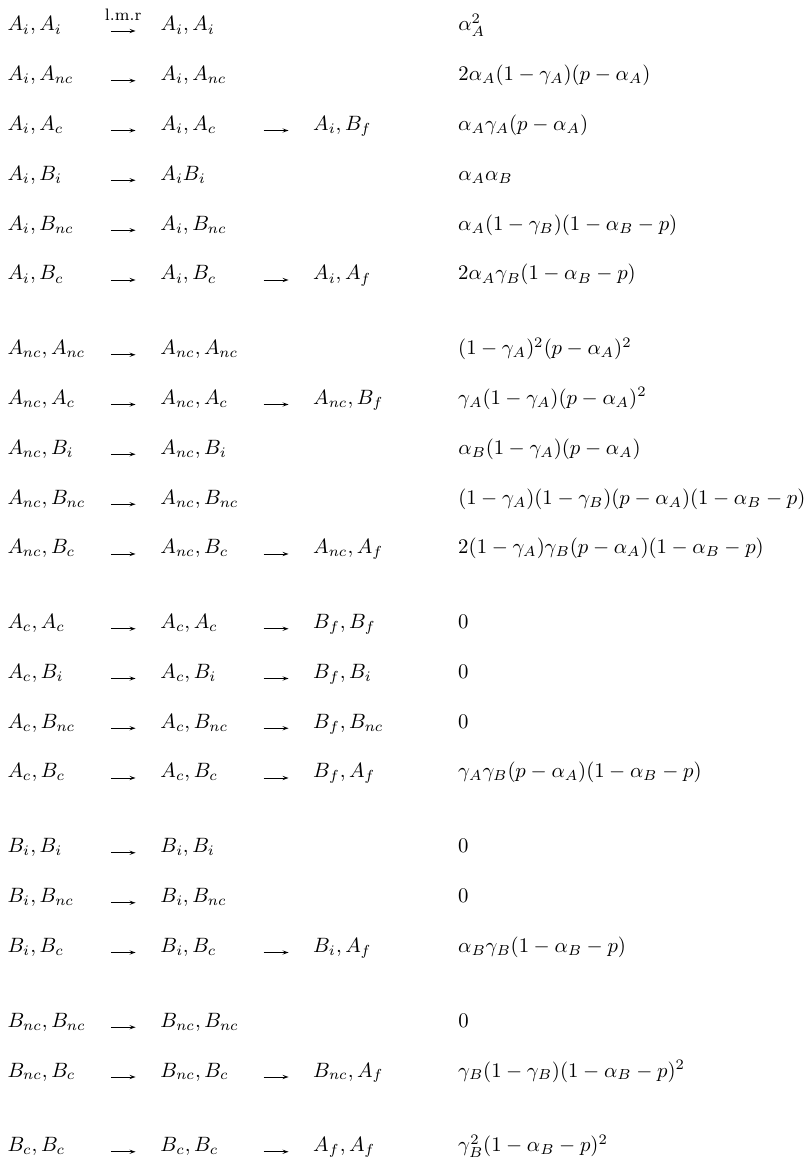}
\end{table}

\newpage

\subsection{Table 3: group size $L=3$}\label{subsection:T3}

\begin{table*}[ht!]
\caption{}\label{cap:tab3}
\begin{tabular}{l}
\includegraphics[width=0.8\textwidth, trim=0cm 2cm 0cm 2cm]{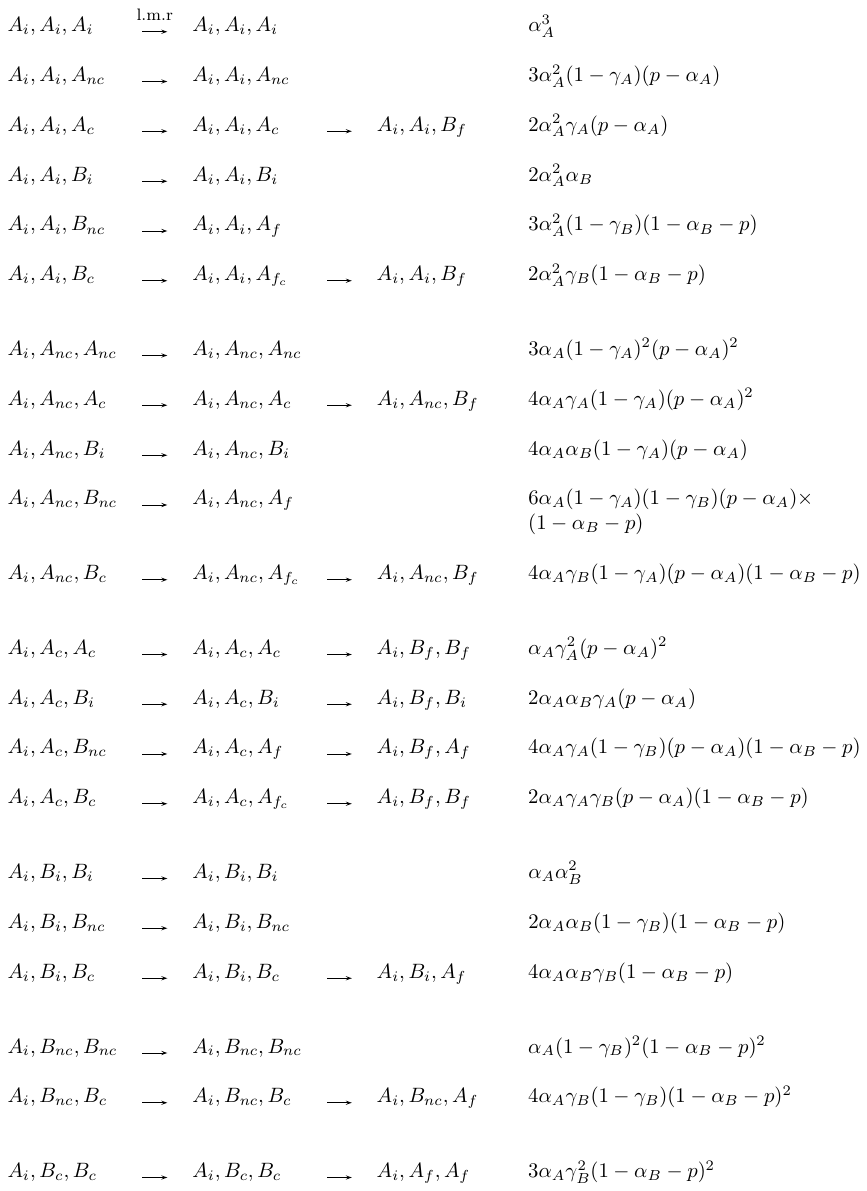}
\end{tabular}
\end{table*}

\newpage

\begin{table*}[ht!]
\begin{tabular}{l}
\includegraphics[width=0.8\textwidth, trim=0cm 2cm 0cm 4cm]{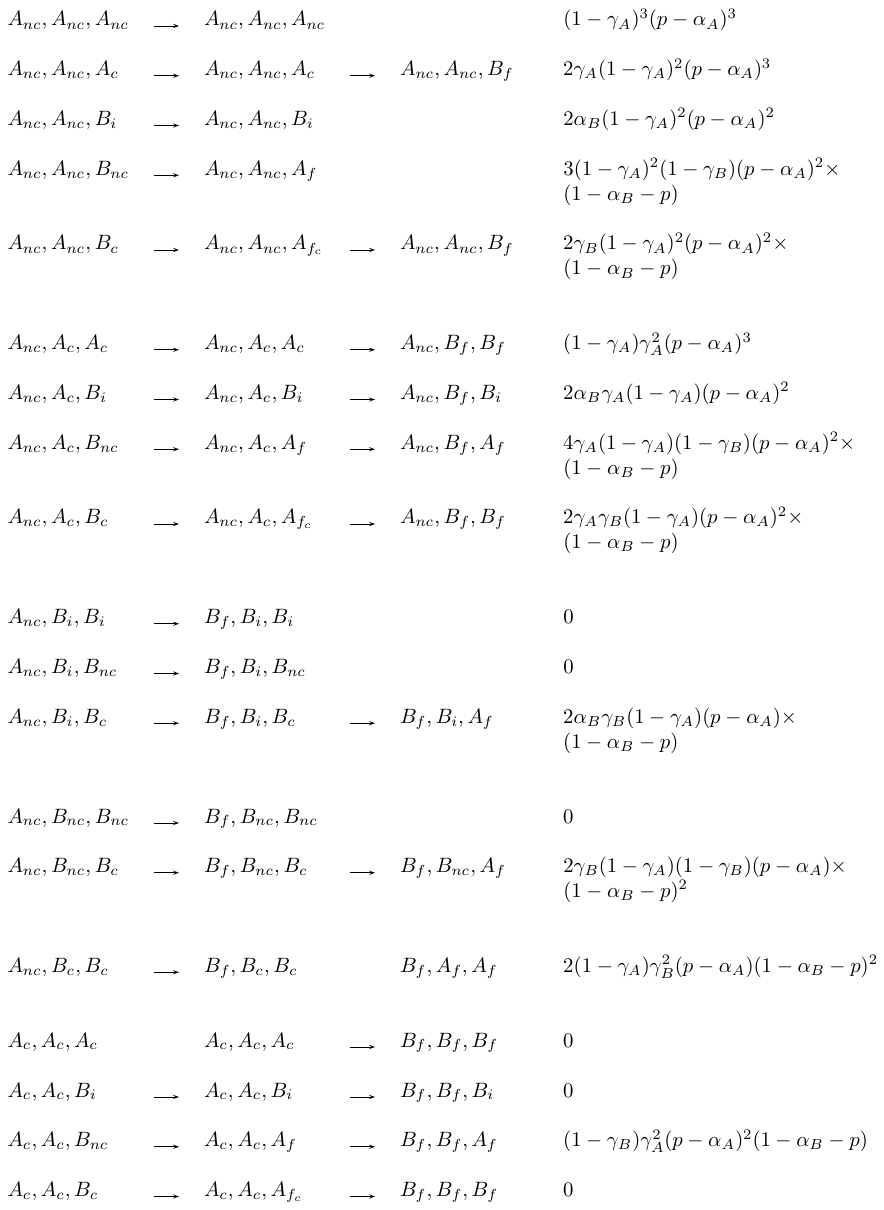}
\end{tabular}
\end{table*}

\newpage

\begin{table}[ht!]
\begin{tabular}{l}
\includegraphics[width=0.8\textwidth, trim=0cm 2cm 0cm 4cm]{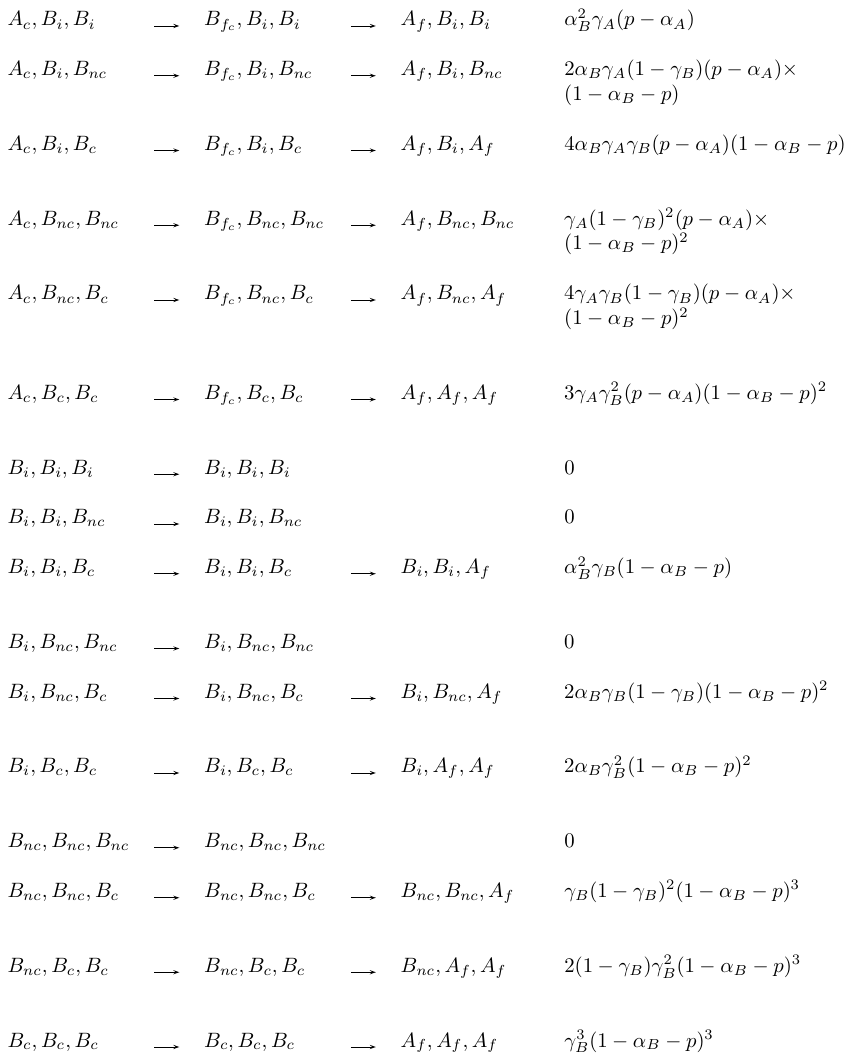}
\end{tabular}
\end{table}

\newpage

\subsection{$L=3$: analysis for the fully symmetric case $\alpha_{A}=\alpha_{B}$ and $\gamma_{A}=\gamma_{B}$}\label{subsection:L=3symm}

\noindent The derivative (with respect to $p$) $f'_{3;\alpha, \alpha;\gamma, \gamma}$ of the function $f_{3;\alpha,\alpha;\gamma,\gamma}$ as given by expression~(\ref{eq:L=3symm}) in the equilibrium $\hat{p}=0.5$ equals $0.5(3-4\alpha)(1-2\gamma)$. There are two additional equilibria 
\begin{equation}
\hat{p}_{3;\alpha, \alpha;\gamma, \gamma}=0.5\frac{1-2\gamma\pm\sqrt{(1-2\gamma)(-2+(3-4\alpha)(1-2\gamma))}}{1-2\gamma}\in[\alpha,1-\alpha]
\end{equation}
\noindent if and only if $0\leq\gamma<\frac{1}{6}$ and $0<\frac{1-6\gamma}{1-2\gamma}-4\alpha$. If the two additional equilibria exist they are symmetrically positioned on opposite sides of $0.5$, and asymptotically stable; the equilibrium $0.5$ then is unstable, with $f'_{3;\alpha, \alpha;\gamma, \gamma}(0.5)>1$.
\\
\noindent For $\alpha$ and $\gamma$ such that $\frac{5}{6}<\gamma\leq1$ and $0<\frac{5-6\gamma}{1-2\gamma}-4\alpha$, the equilibrium $0.5$ also is unstable, with $f'_{3;\alpha, \alpha;\gamma, \gamma}(0.5)<-1$. In this case the dynamics $\overrightarrow{f_{3;\alpha, \alpha;\gamma,\gamma}}$ has two asymptotically stable periodic points $p^{*}_{3;\alpha;\gamma}$ of minimal period 2, symmetrically positioned with respect to $0.5$: 
\begin{equation}
p^{*}_{3;\alpha, \alpha;\gamma, \gamma}=0.5\frac{1-2\gamma\pm\sqrt{(1-2\gamma)(2+(3-4\alpha)(1-2\gamma))}}{1-2\gamma}\in[\alpha,1-\alpha].
\end{equation}

\subsection{$L=3$: analysis of the general case}\label{subsection:L=3general} 

\noindent The possible equilibria for $\overrightarrow{f_{3;\alpha_{A},\alpha_{B};\gamma_{A},\gamma_{B}}}$ (in $[\alpha_{A},1-\alpha_{B}]$) follow from solving $f_{3;\alpha_{A},\alpha_{B};\gamma_{A},\gamma_{B}}(p)=p$, under the restriction that $p\in[\alpha_{A},1-\alpha_{B}]$. We distinguish several cases. 
\begin{enumerate}
\item[1.] $(\gamma_{A},\gamma_{B})=(0.5,0.5)$: expression~(\ref{eq:L=3}) equals $f_{3; \alpha_{A},\alpha_{B};0.5, 0.5}(p)=0.5(1+\alpha_{A}-\alpha_{B})$, and allows for a unique stable equilibrium $\hat{p}=0.5(1+\alpha_{A}-\alpha_{B})$, on which opinion $A$ has the majority if and only if $\alpha_{A}>\alpha_{B}$. 
\item[2.] $(\gamma_{A}, \gamma_{B})\neq(0.5,0.5)$, $\gamma_{A}+\gamma_{B}=1$: the function $f_{3;\alpha_A{},\alpha_{B};\gamma_{A},\gamma_{B}}$ is quadratic in $p$. The discriminant 
\\
$D(\alpha_{A}, \alpha_{B}; \gamma_{A},\gamma_{B})$ for the equation $f_{3;\alpha_{A},\alpha_{B};\gamma_{A},\gamma_{B}}(p)-p=0$ equals $4\alpha_{B}^2 + 4\gamma_{A}(1-\alpha_{A}^2-3\alpha_{B}^2)-4\gamma_{A}^2(1-2(\alpha_{A}^2+\alpha_{B}^2))$. 
The opinion dynamics $\overrightarrow{f_{3;\alpha_{A},\alpha_{B};\gamma_{A}, 1 - \gamma_{A}}}$ has a unique equilibrium 
\\
$\hat{p}_{3;\alpha_{A}, \alpha_{B};\gamma_{A}, 1 - \gamma_{A}}=$
\\\\
$\dfrac{1-\gamma_{A}+(1-2\gamma_{A})\alpha_{A}-\sqrt{(1-\gamma_{A})(1-2\gamma_{A})\alpha_{B}^{2}+\gamma_{A}(1-\gamma_{A})-\gamma_{A}(1-2\gamma_{A})\alpha_{A}^{2}}}{(1-2\gamma_{A})(1+\alpha_{A}+\alpha_{B})}$ 
\\\\
in the interval $[\alpha_{A}, 1 - \alpha_{B}]$. The derivative in the equilibrium equals 
\\\\
$1-2\sqrt{\gamma_{A}(1-\gamma_{A})+(1-\gamma_{A})(1-2\gamma_{A})\alpha_{B}^{2}-\gamma_{A}(1-2\gamma_{A})\alpha_{A}^{2}}$.
\item[3.]$\gamma_{A}+\gamma_{B}\neq1$. The expression $f_{3;\alpha_{A},\alpha_{B};\gamma_{A},\gamma_{B}}(p)-p=0$ for determining the equilibria now is 
\begin{equation}\label{eq:g}
\begin{tabular}{l}
$f_{3;\alpha_{A},\alpha_{B};\gamma_{A},\gamma_{B}}(p)-p=$
\\\\
$\alpha_{A}(1 - \gamma_{A}) + (1 - \alpha_{B})\gamma_{B} - \Big(1 + 2\alpha_{A}(1 - 2 \gamma_{A}) - \gamma_{A} + \gamma_{B}\Big)p_{t} +$
\\\\
$\Big(3 + \alpha_{A}(1-2\gamma_{A})-\alpha_{B}(1-2\gamma_{B})-4\gamma_{A}-2\gamma_{B}\Big)p_{t}^2 -2\Big(1-\gamma_{A}-\gamma_{B}\Big)p_{t}^3=0.$
\end{tabular}
\end{equation}
Its discriminant is 
$
D(\alpha_{A},\alpha_{B};\gamma_{A},\gamma_{B})=\Big(\frac{1}{2}q_{1}(\alpha_{A},\alpha_{B};\gamma_{A},\gamma_{B})\Big)^{2}+\Big(\frac{1}{3}q_{2}(\alpha_{A},\alpha_{B};\gamma_{A},\gamma_{B})\Big)^{3}
$ with 
\\
\noindent $c_{0}(\alpha_{A},\alpha_{B};\gamma_{A},\gamma_{B})=\alpha_{A}(1-\gamma_{A})+(1-\alpha_{B})\gamma_{B}$,
\\
\noindent $c_{1}(\alpha_{A},\alpha_{B};\gamma_{A},\gamma_{B})=-(1+2\alpha_{A}(1-2\gamma_{A})-\gamma_{A}+\gamma_{B})$,
\\
\noindent $c_{2}(\alpha_{A},\alpha_{B};\gamma_{A},\gamma_{B})=3+\alpha_{A}(1-2\gamma_{A})-\alpha_{B}(1-2\gamma_{B})-4\gamma_{A}-2\gamma_{B}$,
\\
\noindent $c_{3}(\alpha_{A},\alpha_{B};\gamma_{A},\gamma_{B})=-2(1-(\gamma_{A}+\gamma_{B}))$,
\\
\noindent and 
\\
$q_{1}(\alpha_{A},\alpha_{B};\gamma_{A},\gamma_{B})=
\dfrac{2}{27}\Big(\dfrac{c_{2}(\alpha_{A},\alpha_{B};\gamma_{A},\gamma_{B})}{c_{3}(\alpha_{A},\alpha_{B};\gamma_{A},\gamma_{B})}\Big)^{3}-\dfrac{1}{3}\dfrac{c_{2}(\alpha_{A},\alpha_{B};\gamma_{A},\gamma_{B})}{c_{3}(\alpha_{A},\alpha_{B};\gamma_{A},\gamma_{B})}\dfrac{c_{1}(\alpha_{A},\alpha_{B};\gamma_{A},\gamma_{B})}{c_{3}(\alpha_{A},\alpha_{B};\gamma_{A},\gamma_{B})}+\dfrac{c_{0}(\alpha_{A},\alpha_{B};\gamma_{A},\gamma_{B})}{c_{3}(\alpha_{A},\alpha_{B};\gamma_{A},\gamma_{B})}$,
\\\\
$q_{2}(\alpha_{A},\alpha_{B};\gamma_{A},\gamma_{B})=-\dfrac{1}{3}\Big(\dfrac{c_{2}(\alpha_{A},\alpha_{B};\gamma_{A},\gamma_{B})}{c_{3}(\alpha_{A},\alpha_{B};\gamma_{A},\gamma_{B})}\Big)^{2}+\dfrac{c_{1}(\alpha_{A},\alpha_{B};\gamma_{A},\gamma_{B})}{c_{3}(\alpha_{A},\alpha_{B};\gamma_{A},\gamma_{B})}$.
\\\\
Figure~\ref{cap:p3selectionalt} shows a selection of signplots of the discriminant for this case. In addition the results of the analysis for combinations $(\alpha, \alpha; \gamma, \gamma)$ is included. 
\end{enumerate}

\newpage

\noindent Conflict of interest disclosure:
\\
The authors declare that there is no conflict of interest regarding the publication of this article.


\newpage

\begin{thebibliography}{999}

\bibitem{strike} S. Galam, Y. Gefen and Y. Shapir, Sociophysics: A new approach of sociological collective behaviour. 1 Mean-behaviour of a strike, Math. J. Sociology 9 (1982) 1-13

\bibitem{book} S. Galam, Sociophysics: A Physicist's Modeling of Psycho-political Phenomena, Springer (2012)

\bibitem{perony} N. Perony, R. Pfitzner, I. Scholtes, C. J. Tessone, F. Schweitzer, Enhancing consensus under opinion bias by means of hierarchical decision making, Advances in Complex Systems 16 (06) (2013) 1350020

\bibitem{schweitzer} F. Schweitzer, Sociophysics, Physics Today 71, 2, 40 (2018); doi: 10.1063/PT.3.3845

\bibitem{castellano}  C. Castellano, S. Fortunato and C. Vittorio Loreto,  Statistical Physics of Social Dynamics, Rev. Mod. Phys. 81 (2009)  591-646

\bibitem{mino1} S. Galam,  Minority Opinion Spreading in Random Geometry, Eur. Phys. J. B 25, Rapid Note (2002) 403

\bibitem{mino2} S. Galam, The dynamics of minority opinion in democratic debate, Phys. A 336 (2004) 56

\bibitem{uni} S. Galam, Local dynamics vs. social mechanisms: A unifying frame, Europhys. Lett. 70 (2005) 705-711 

\bibitem{three} S. Gekle, L. Peliti, and S. Galam, Opinion dynamics in a three-choice system, Eur. Phys. J. B 45 (2005) 569-575 

\bibitem{wonc} S. Galam, S. Wonczak, Dictatorship from majority rule voting, Eur. Phys. J. B 18 (2000) 183-186

\bibitem{contra1} S. Galam, Contrarian deterministic effects on opinion dynamics: the hung elections scenario, Physica A (2004) 333, 453-460 

\bibitem{contra2} S. Galam, From 2000 Bush-Gore to 2006 Italian elections: voting at fifty-fifty and the contrarian effect, Qual. Quant, 41 (2007) 579-589

\bibitem{schneider} J. J. Schneider,  The influence of contrarians and opportunists on the stability of a democracy in the Sznajd model, International Journal of Modern Physics C 15 (2004) 659-674

\bibitem{stauf} D. Stauffer, J. S. S\'a Martins, Simulation of Galam's contrarian opinions on percolative lattices, Physica A 334, (2004) 558-565 

\bibitem{wio1} H. S. Wio, M. S. de la Lama, and J. M. L\'opez, Contrarian-like behavior and system size stochastic resonance in an opinion spreading model, Physica A 371 (2006) 108-111

\bibitem{wio2} M. S. de la Lama, J. M. L\'opez, and H. S. Wio, Spontaneous emergence of contrarian-like behaviour in an opinion spreading model, Europhys. Lett. 72 (2005) 851-857 

\bibitem{mobi1} M. Mobilia, A. Petersen, and S. Redner, On the role of zealotry in the voter model, J. Stat. Mech. Volume 2007 (2007) P08029 

\bibitem{chiche} C. Borghesi, J. Chiche and J. P. Nadal, Between order and disorder: a `weak law' on recent electoral behavior among urban voters? PLoS One Volume 7 (2012) 0039916 

\bibitem{masuda} N. Masuda, Voter models with contrarian agents, Phys. Rev. E 88 (2013) 052803

\bibitem{kasia1} K. Sznajd-Weron, J. Szwabinski and R. Weron, Is the Person-Situation Debate Important for Agent-Based Modeling and Vice-Versa?, Plos One 9 (2014) e0112203

\bibitem{weis} G. Weisbuch, From Anti-Conformism to Extremism, JASSS18 (2015) 1

\bibitem{chatt} S. Mukherjee and A. Chatterjee, Disorder-induced phase transition in an opinion dynamics model: Results in two and three dimensions, Phys. Rev. E 94 (2016) 062317

\bibitem{taksu} T. Cheon and J. Morimoto, Balancer effects in opinion dynamics, Physics Letters A 380 (2016) 429-434

\bibitem{nuno1}  J. P. Gambaro and N. Crokidakis, The influence of contrarians in the dynamics of opinion formation, Physica A 486 (2017) 465-472

\bibitem{timpa} A. M. Timpanaro, Diversity and Disorder in the Voter Model with Delays, https://arxiv.org/abs/1708.08756v2

\bibitem{mas1} M. M\"as, A, Flache and D. Helbing, Individualization as Driving Force of Clustering Phenomena in Humans, PLoS Comp. Biol. 6 (10): e1000959. https://doi.org/10.1371/journal.pcbi.1000959 (2010)

\bibitem{mas2} M. M\"as, A. Flache and J.A. Kitts, Cultural Integration and Differentiation in Groups and Organizations. In: V. Dignum, F. Dignum (eds.), Perspectives on Culture and Agent-based Simulations. Studies in the Philosophy of Sociality, vol 3. Springer International Publishing Switzerland, DOI $10.1007/978-3-319-01952-9_5$ (2014)

\bibitem{moham} A. Mohammadinejad, R. Farahbakhsh, N. Crespi, Consensus Opinion Model in Online Social Networks Based on Influential Users, IEEE Access Volume 7, IEEE, DOI: 10.1109/ACCESS.2019.2894954 (2019) 28436-28451

\bibitem{mosco} S. Galam and S. Moscovici, Towards a theory of collective phenomena: Consensus and attitude changes in groups, European Journal of Social Psychology 21 (1991) 49-74

\bibitem{inflex1} S. Galam and F. Jacobs, The role of inflexible minorities in the breaking of democratic opinion dynamics, Physica A, 381 (2007) 366-376 

\bibitem{public} S. Galam, Public debates driven by incomplete scientific data: The cases of evolution theory, global warming and H1N1 pandemic influenza, Physica A 389 (2010) 3619-3631

\bibitem{iglesia} S. Goncalves, M. F. Laguna and J. R. Iglesias, Why, when, and how fast innovations are adopted, Eur. Phys. J. B 85 (2012) 192

\bibitem{martins} A. C. R. Martins and S. Galam, Building up of individual inflexibility in opinion dynamics, Phys. Rev. E 87 (2013) 042807 

\bibitem{mobi2} M. Mobilia, Nonlinear q-voter model with inflexible zealots, Phys. Rev. E 92 (2015) 012803 

\bibitem{bolek1} W. Pickering, B. K. Szymanski and C. Lim, Analysis of the high dimensional naming game with committed minorities, Phys. Rev. E (2016) 93:0523112

\bibitem{rand} K. Burghardt, W. Rand and M. Girvan, Competing opinions and stubborness: Connecting models to data, Phys. Rev. E 93  (2016) 032305 

\bibitem{bollen} N. Rodriguez, J. Bollen and Y. Y. Ahn, Collective Dynamics of Belief Evolution under Cognitive Coherence and Social Conformity, PLoS One Volume 11 (2016) 0165910 

\bibitem{latora} F. Battiston, A. Cairoli, V. Nicosia, A. Baule and V. Latora, Interplay between consensus and coherence in a model of interacting opinions, Physica D 323 (2016) 12-19

\bibitem{anten} A. M. Calv\~ao, M. Ramos and C. Anteneodo, Role of the plurality rule in multiple choices,  J. Stat. Mech. Volume 2016 (2016) 023405

\bibitem{sob} P. Sobkowicz, Social Simulation at the Ethical Crossroads, Science and Engineering Ethics 25 (1) (2019) 143-157

\bibitem{kasia2} A. Jedrzejewski and K. Sznajd-Weron, Person-Situation Debate Revisited: Phase Transitions with Quenched and Annealed Disorders, Entropy 19 (2017) 415

\bibitem{nuno2} M. A. Pires and N. Crokidakis, Dynamics of epidemic spreading with vaccination: Impact of social pressure and engagement, Physica A 467 (2017) 167-179

\bibitem{lee} E. Lee, P. Holme and S. H. Lee, Modeling the dynamics of dissent, Physica A 486 (2017) 262-272

\bibitem{taksu2} S. Galam and Cheon, T., Tipping Point Dynamics: A Universal Formula, arXiv: 1901.09622v1 [physics.soc-ph] (2019)

\end{thebibliography}
\end{document}